\documentclass[aps,prl,twocolumn,groupedaddress,longbibliography,showkeys,showpacs]{revtex4-1}

\usepackage{subcaption}% preamble
\usepackage{graphicx}% Include figure files
\usepackage{bm}% bold math
\usepackage{amssymb,amsfonts}
\usepackage{amsmath}
\usepackage{natbib}
\usepackage{graphicx}
\usepackage[usenames,dvipsnames]{xcolor}
\usepackage{soul}
\usepackage{url}
\begin{document}

\title{In Response to COVID-19: Configuration Model of the Epidemic Spreading}

%\title{Statistical Description of Complex Scale-free Networks}

\author{Alexander I. Nesterov}
   \email{nesterov@cencar.udg.mx}
\affiliation{Departamento de F{\'\i}sica, CUCEI, Universidad de Guadalajara,
 Guadalajara, CP 44420, Jalisco, M\'exico}

\author{Pablo H\'ector Mata Villafuerte}
   \email{themata@hotmail.com}
\affiliation{Departamento de F{\'\i}sica, CUCEI, Universidad de Guadalajara,
 Guadalajara, CP 44420, Jalisco, M\'exico}

\author{Gennady P  Berman}
 \email{gpb@lanl.gov}
\affiliation{Theoretical Division, T-4, Los Alamos National Laboratory,  Los Alamos, NM 87545, USA}

\date{\today}

\begin{abstract}
A configuration model for epidemic spread, based on a scale-free network, is introduced.  We obtain the analytical solutions describing both unstable and stable dynamics of the epidemic spreading,  and demonstrate how these regimes can interchange during the epidemic. We apply the model to the COVID-19 case and demonstrate the predictive features of our model. 

\end{abstract}

%\pacs{89.75.Hc, 89.20.Hh,02.50.-r, 05.30.-d}

\preprint{ LA-UR-20-24290}

\maketitle
\section{Introduction}

According to the World Health Organization (WHO), the COVID-19 epidemic started in the 
Chinese city of Wuhan on December 31, 2019. The speed of the epidemic spread is very high, 
and on March 11, 2020, the WHO declared the COVID-19 outbreak a pandemic. Now, 
more than 200 countries in the world are affected by the coronavirus epidemic. COVID-19 
presents an immense challenge for the scientific community. The advances in 
understanding of the disease's properties and the mechanisms of its spreading will allow scientists to better predict the dynamics of the epidemic and the characteristic time of recovery. They will also help to make the right decisions for overcoming the pandemic \cite{AD,IBMRF,CS}. 

Network science has contributed to diverse fields in both the natural and human sciences.  We hope that,  due to its intrinsic interdisciplinary nature, the network approach will be useful for understanding the dynamics of the COVID-19 pandemic. One of the reasons for employing network science in the study of this phenomenon is the analogy between the spread of information in social networks and the spread of disease by contact between individuals. For instance, the spread of news, rumors, or gossip through a population has features in common with disease spread. The ideas and models for dissemination of information in networks can help us to better understand the propagation of disease \cite{BB1,NMEJ2,MSRT,LAMR,BAL,DSMF,BABM,MN2018}. 

A critical issue in mathematical modeling of the epidemic is a choice of the adequate 
mathematical tools and a suitable network model, reflecting the features of real 
networks. The fascinating discovery of contemporary network science is the universality of network  topology. Many real networks, such as social networks, airline networks, the 
World Wide Web, computer networks, the Internet, urban networks, and others, exhibit scale-invariance, so they can be treated as {\em scale-free networks} \cite{GCal}. Their scale-free nature, and other features, make them an excellent candidate for the mathematical modeling of the epidemic spreading
\cite{MRLA,GCal,BAL1,MN2018,GMNM,ARB}.  

The idea that a statistical approach is adequate to study complex networks is a natural one since networks are large complex systems, and a deterministic approach cannot describe their collective behavior. Nowadays, the methods of statistical mechanics have become powerful tools for the study and explanation of real-world network properties \cite{MEJN1,RPMR,ARB,MN2018,PJNM,CDLM,CGST,BAL1,BAL}.

In this paper we study the connection between network structure and the epidemic spreading 
using statistical physics methods, and propose a mathematical model based on scale-free 
networks which allows us to extrapolate the dynamics of an epidemic. We obtain the analytical 
solutions that describe both unstable and stable dynamics, and demonstrate how these regimes 
interchange during the process of evolution of the epidemic. We apply the model to the 
COVID-19 case and demonstrate the predictive features of our approach. To conclude, we 
discuss the existing issues of our approach and possible future developments. In the 
Supplemental Material (SM), we present the results of the application of our approach for 
COVID-19.

\section{Statistical description of networks} 

A network is a set of $N$ nodes (or vertices) connected by $L$ links (or edges). One can describe the network by an adjacency matrix, $a_{ij}$, where each existing (or non-existing) link between pairs of nodes ($ij$) is indicated by 1 or 0 in the $i,j$ entry. Individual nodes possess local properties such as node degree, $k_i = \sum_j a_{ij}$, and clustering coefficient \cite{WDSS,NMSW,BSLV,ARB}. The whole network can be described quantitatively by its degree distribution, $P_k$, and its connectivity. The latter is characterized by the connection probability $p_{ij}$, i.e., the probability that a pair of nodes $(ij)$ is connected.  

The most general statistical description of an undirected network in equilibrium, with a fixed number of vertices, $N$, and a varying number of links, $L$, is given by the grand canonical ensemble. The probability of obtaining a graph, $A$, with the energy, $E_A$, and the number of links, $L_{A}=\sum_{i j} a_{i j}$, can be written as
\cite{PJNM,CDLM,CDAS,CGST,WJWR},
\begin{align}
P_{A}=\frac{1}{\mathcal{Z}} \exp \big (\beta(\mu L_{A}-E_{A})\big),
\end{align}
where  $\beta =1/T$, with $T$ being the network temperature, and $\mu$ is the chemical potential. The partition function reads
\begin{align}
\mathcal{Z} =\sum_{A} \exp \big (\beta(\mu L_{A}-E_{A})\big).
\end{align}
The temperature is a parameter that controls clustering in the 
network, and the chemical potential controls the link density and 
the connection probability. 
 
Let us assign to each edge $\langle i,j \rangle$ the link energy, $\varepsilon_{ij}$. Then, the energy of the graph can be written as $E_{A} =\sum_{i<j} \varepsilon_{i j} a_{i j}$, and the partition function and the graph probability are given by \cite{CDAS}
\begin{align} \label{Z}
	\mathcal{Z}=&\prod_{i<j}\big (1+e^{\beta\left(\mu-\varepsilon_{i j}\right)  }\big ), 
	\\
P_{A}=&\prod_{i<j} p_{i j}^{a_{i j}}\left(1-p_{i j}\right)^{1-a_{i j}}.
\end{align}
Here, $p_{ij}$ is the connection probability between nodes $i$ and $j$, and has the usual form of the Fermi-Dirac distribution \cite{PJNM,CDAS,CDLM},
\begin{align}
	p_{i j}=\frac{1}{e^{\beta \left(\varepsilon_{i j}-\mu\right) }+1}.
	\label{EqP}
\end{align}

To obtain the thermodynamical potentials, we use the partition function and the usual definitions from statistical physics. For instance, the Landau free energy (grand canonical potential) is given by $\Omega= - \beta^{-1} \ln\mathcal Z$. We can recover the Helmholtz free energy, $F$, and internal energy, $E$, using the relations $F = \Omega + \mu L$ and  $ E = F+  \beta {\partial F}/{\partial \beta}$. Finally, having the Landau free energy, one can obtain the equation of state by writing $L = -\partial \Omega/\partial \mu$, where $L$ is the expected number of links.

Let us assume that the link energy in Eq. \eqref{EqP} is the sum of the contributions from each node, $\varepsilon_{ij} =\epsilon_i + \epsilon_j$. Then the energy of the graph, $ A$, can be written as $E_A =\sum_i \epsilon_i k_i$, where $ k_i = \sum_j a_{ij}$.  Thus, all graphs $A$ with equal degrees have the same energy, $E_A$. This model is a particular case of the so-called {\em Configuration Model}, i.e., a model of a random network with given degree sequence, rather than with a degree distribution \cite{PJNM,CDAS}.

We will focus now on the configuration model based on scale-free networks. A scale-free network is characterized by a power-law degree distribution,  $\rho(k_i)\sim (\gamma-1) k_i^{-\gamma}$, where $1\leq k_i \leq k_0$, and the exponent of the distribution is $\gamma > 1$. Let us assign to each node a hidden variable $\epsilon_i$ as follows: $ k_i =  e^{-\beta_c \epsilon_i}$, where $\beta_c $ is a constant with dimension of inverse temperature. As above, the link energy of the edge $\langle i j \rangle$ is given by $\varepsilon_{ij} = \epsilon_{i}  + \epsilon_{j} $. The quantities $\epsilon_i$  are distributed according to $\rho(\epsilon_i)\sim \beta_c(\gamma -1)e^{\beta_c(\gamma- 1)\epsilon_i}$, where $0 \leq \epsilon_i \leq \epsilon_0$ and $\epsilon_0=  T_c \ln k_0$ \cite{CDAS}. Due to the scale invariance of the network's topological properties, the parameter $\beta_c$ is a dummy parameter and can be chosen arbitrarily.

In the continuous limit, we have $\rho(\epsilon_i ) \rightarrow \rho(\epsilon )$, where $\rho(\epsilon) = C e^{\alpha\beta \epsilon}$ and $\alpha = \beta_c(\gamma -1)/\beta$. The constant $C$ is defined by the normalization condition $\int\limits_0^{\epsilon_0}\rho(\epsilon) d\epsilon =1$. This yields
\begin{align}
    \rho(\epsilon) =\frac{\alpha \beta e^{ \alpha \beta(\epsilon - \epsilon_0/2)}}{2\sinh(\alpha \beta \epsilon_0/2)} .
\end{align}

\section{ Configuration model of the epidemic spreading} 

In our model, each node in the network is identified with an individual. The probability of an individual being infected by an infected individual depends on the connection degree, susceptibility of the population, and the connection probability. Since the transmission of the epidemic depends on the edge between two nodes being occupied or not, the connectivity is essential for the epidemic spread. We assume that the disease spreading does not take the network out of the thermodynamic equilibrium.

The assumption that the disease may propagate only along the links in the network is usual for the models dealing with the spread of epidemics on networks (see, i. e. \cite{MN2018}).  We assume that the probability of contact between pairs of individuals leading to disease can vary, just as in  Ref. \cite{NMEJ2}. Thus, some pairs can have a higher probability of disease transmission than others. However, the ``connection'' (or existing link) does not guarantee the disease-causing contact. Therefore, we distinguish infective individuals, who may infect others, and susceptible individuals, who can be infected. Note that there are a tiny number of infectious individuals at the beginning of a disease outbreak. Since the transmission of infection is a stochastic process, depending on the pattern of contacts between individuals, a description should take this arrangement into account \cite{FBCC}.
 
Consider a pair of nodes, $(i,j)$, where one node, $i$, is infective, and the other one, $j$, is susceptible to infection.  Suppose that the transmission rate of disease between nodes is $r_{ij} $. Then, the average number of infected nodes is given by $N_c = \sum_{\langle i j \rangle} p_{ij} r_{ij}$. In the continuous limit this yields $N_c = \iint p(\epsilon',\epsilon'') r(\epsilon',\epsilon'')\rho(\epsilon')\rho(\epsilon'') d\epsilon' d\epsilon''$.

We divide the population into classes with a fixed link energy of pairs, $\varepsilon = \epsilon_i + \epsilon_j$, and consider only the pairs of individuals in which one is infective, and the other one is susceptible. We assume that the transmission rate of disease depends only on the link energy, and present this rate in the simplest form: $r_{ij} = C \delta( \epsilon_i+\epsilon_j- \varepsilon )$, where $C$ is a constant.  Now, the average number of infectious nodes, $N_c (\varepsilon)$,  with a given energy, $\varepsilon$, is given by
\begin{align}
N_c(\varepsilon)     =  C \iint_0^{\epsilon_0}\frac{\rho(\epsilon')\rho(\epsilon'') \delta( 
\epsilon'+\epsilon''- \varepsilon )d\epsilon' d\epsilon''}{e^{\beta (\epsilon' + \epsilon''-\mu) 
}+1} .
\label{Eq.8a}
\end{align}
Performing the integration, we obtain
\begin{align}
N_c(\varepsilon)     = \frac{A e^{\alpha \beta (\varepsilon-\mu) } }{e^{\beta (\varepsilon-\mu) }+1}.
\end{align}
Here the constant $A$ accumulates all constants that the integral \eqref{Eq.8a} includes.

The number of infected nodes, $\Delta N_t$, inside the interval $\beta \Delta \varepsilon$ can be written as $\Delta N_t     = N_c \beta \Delta \varepsilon$. (The inverse temperature, $\beta$, is introduced to keep the right dimensionality.) Performing the integral, $N_t  =\beta \int N_c(\varepsilon) d\varepsilon$, we obtain
\begin{align}
N_t  = A e^{\alpha \beta (\varepsilon- \mu)} \Phi\big (- e^{\beta (\varepsilon -\mu)} ,1, \alpha\big ) + B,
\label{Eq.9}
\end{align}
where $B$ is a constant of integration, and $\Phi(z,s,a)$ is the Lerch Transcendent \cite{NIST}. 

Now we consider the epidemic spreading. We assume that the link energy is an increasing function of time. Then from Eq. \eqref{Eq.9} it follows
\begin{align}
\frac{dN_t}{d t} =  N_i ,
\label{Eq.10}
\end{align}
where $N_i=\kappa N_c$ is the number of infectious individuals ({\em incidence}) per unit of time (usually daily), and  $\kappa(t) = \beta\dot \varepsilon (t)$ is the transmission rate of the epidemic disease. The rate, $\kappa$, determines the expected number of people that an infected person infects per time, as the SIR model defines it \cite{MM1, FBCC, MRA}. 
 
To understand the impact of the parameter $\alpha$, consider the asymptotic solution of Eq. \eqref{Eq.10}. 
In the limit of $\varepsilon(t)\gg \mu$,  we  have
$N_i  \sim  e^{ (\alpha -1)\beta(\varepsilon (t) -\mu)}$. Substituting $N_i$ in Eq. \eqref{Eq.10}, we obtain
\begin{align}
N_t \sim \left \{\begin{array}{lc} 1-
  e^{ (\alpha -1)\beta\varepsilon (t)} &  \alpha  <1\\
  \beta \varepsilon (t), & \alpha =1 \\
  e^{ (\alpha -1)\beta\varepsilon (t)}, &  \alpha  >1
\end{array}
\right .	
\end{align}
It follows that, for $\alpha \geq 1$, the total number of infected nodes $N_t \rightarrow\infty$ when $t \rightarrow \infty$. For any choice of $\alpha $, the obvious condition $N_t \leq N$ should be imposed. 

Thus, $\alpha$ is a crucial parameter for the epidemic spreading, a {\em critical spreading 
parameter}. It determines the threshold of the outbreak. When $\alpha  \geq 1$, the system has 
lost stability, and the process of the epidemic spreading becomes uncontrollable. 

\subsection{ Dynamical system beyond the network model} 

 Here we present our configuration network model as the system of ordinary differential equations, and compare it with the well-known SIR model \cite{BFR, FBCC}.

From Eq. \eqref{Eq.10} it follows that $N_c$ satisfies the following differential equation:
\begin{align}
        \frac{ dN_c}{d t}=   a  N_c \Big (1 - \frac{N_c }{K}\Big),
    \label{Eq10b}
\end{align}
where  $a = \alpha \kappa$, and $K =\alpha A  e^{(\alpha -1)\beta (\varepsilon -   \mu) }$ is the {\em carrying capacity}. As one can see,  Eq. \eqref{Eq10b} is just the modified logistic equation, widely used for description of population growth (see, i.e., \cite{BFR, FBCC}).  When $\kappa = \rm const$  and $\alpha =1$, we have $K= \rm const$, and  \eqref{Eq10b} becomes the standard logistic equation.

Adding to Eqs. \eqref{Eq.10}  and \eqref{Eq10b} the equation for the carrying capacity, we obtain the decoupled system of ordinary differential equations that describe our configuration model:
\begin{align} \label{Eq14a}
    \frac{ dN_t}{d \tau} =& N_c, \\
\frac{ dK}{d \tau} = &   (\alpha -1 ) K , 
    \label{Eq14c} \\
        \frac{ dN_c}{d\tau}= &  \alpha  N_c \Big (1 - \frac{N_c }{K}\Big),
    \label{Eq14b}
\end{align}
where $\tau = \int \kappa(t) dt$ is a dimensionless time. The system of Eqs. (13)-(15) is completely equivalent to Eqs. (9) and (10). The parameter $\alpha$ now is interpreted as the intrinsic growth rate. In Eqs. \eqref{Eq14a} -- \eqref{Eq14b}, $N_t$ is the total number of infected individuals, $N_c$ is the number of new infective individuals, and $K$ denotes the carrying capacity of the infective population. The  number of infected individuals per unit of time can be obtained from the relation $N_i = \kappa N_c$. The solutions of Eqs. \eqref{Eq14a} -- \eqref{Eq14b} should be considered for $N_t \leq N$, where $N$ is the total population.

The most commonly used models for epidemic transmission are the 
Susceptible-Infected-Susceptible (SIS) and the Susceptible-Infected-Removed (SIR). While in the 
SIS model recovered individuals could again be infected, the SIR model assumes that those 
recovered from the disease have immunity, and therefore each individual can only be infected 
once \cite{MM1, FBCC, MRA}. 

The SIR model divides the population into the following three classes.  Susceptible ($S$): individuals in the susceptible state may be infected when they encounter an infected individual. Infectious ($I$): the individuals with the disease, meaning they have the disease and can spread disease by infecting others. Removed ($R$): those who have recovered from the disease (or deceased) or have immunity. The sum of the three numbers is a constant, $S+ I + R =N$, where $N$ is the population. The SIR system is described by the following system of ordinary differential equations \cite{BFR, FBCC}:
\begin{align}\label{SIR1}
\frac{d S}{d t} &=-\frac{\nu I S}{N} ,\\
\frac{d I}{d t} &=\frac{\nu I S}{N}-\lambda I ,
\label{SIR2}\\
\frac{d R}{d t} &=\lambda I.
\label{SIR3}
\end{align}
Here ${\nu I S}/{N}$ is the number of new individuals infected per unit of time, with $\nu$ being the infection rate. 
\begin{figure*}[tbh]
   % \centering
    \begin{subfigure}[t]{0.3\textwidth}
        \centering
        \includegraphics[height=1.6in]{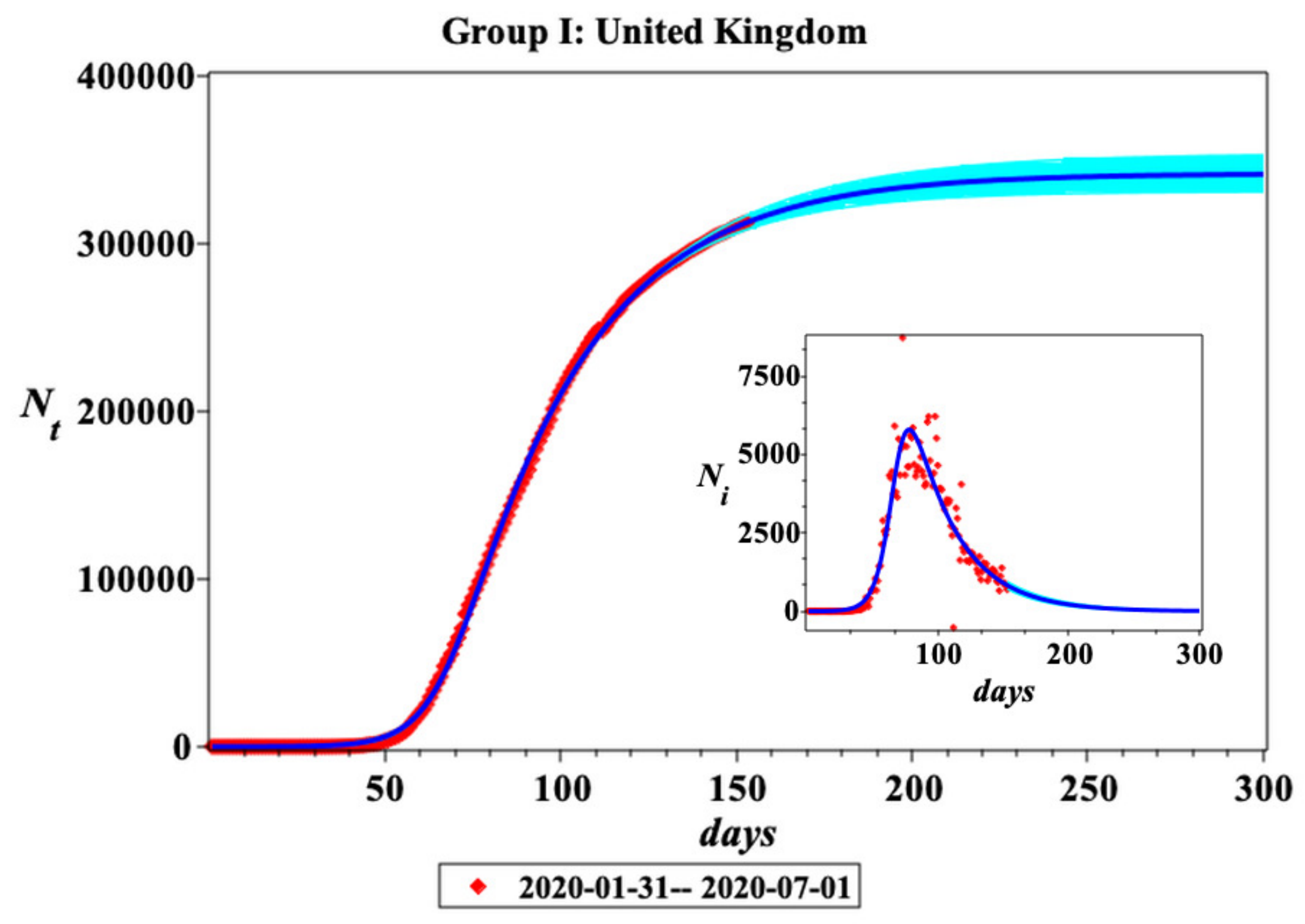} 
    \end{subfigure}
        \quad \quad
    \begin{subfigure}[t]{0.3\textwidth}
        %\centering
        \includegraphics[height=1.65in]{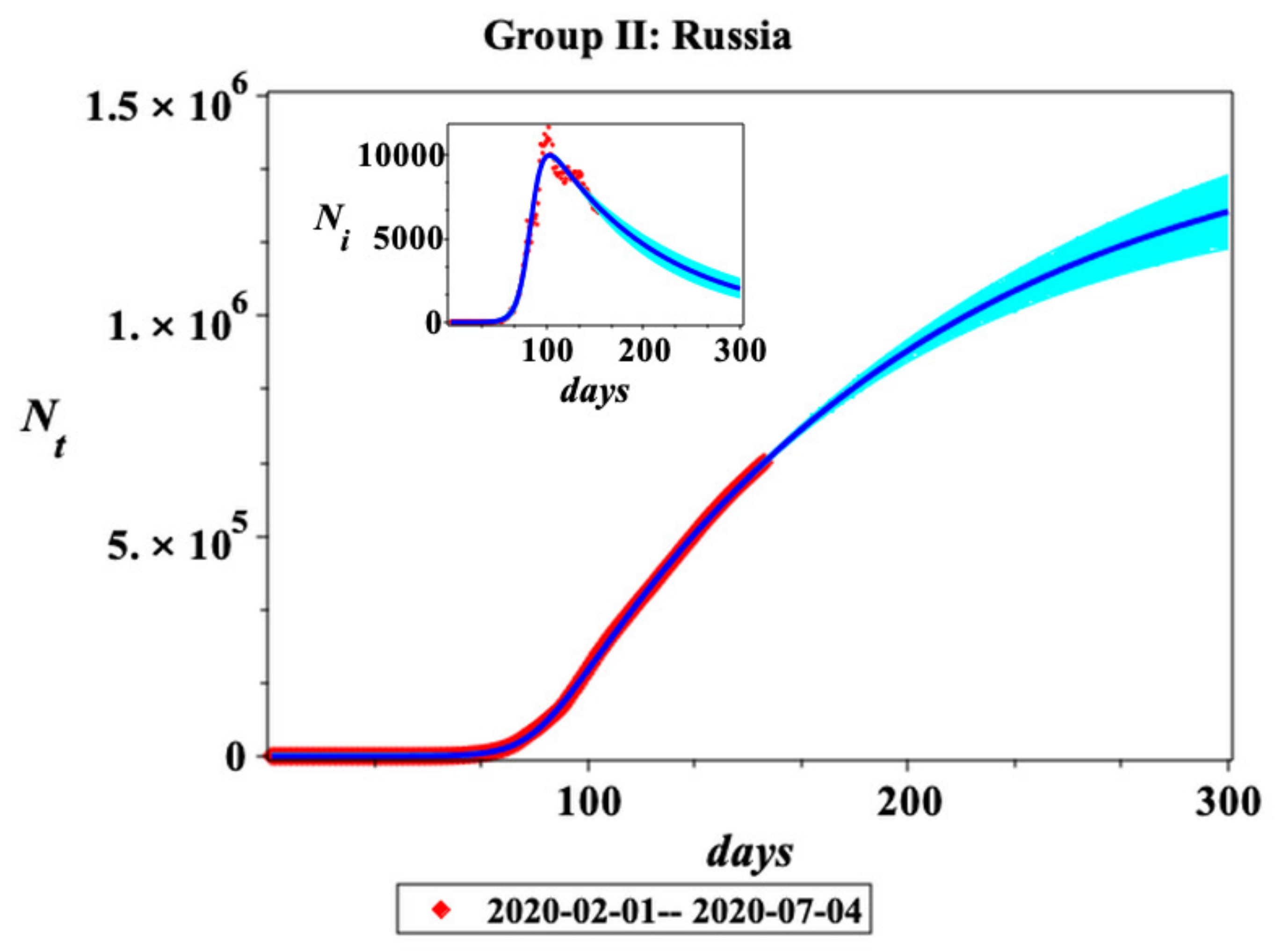} 
    \end{subfigure}
    \quad
    \begin{subfigure}[t]{0.3\textwidth}
        %\centering
        \includegraphics[height=1.68in]{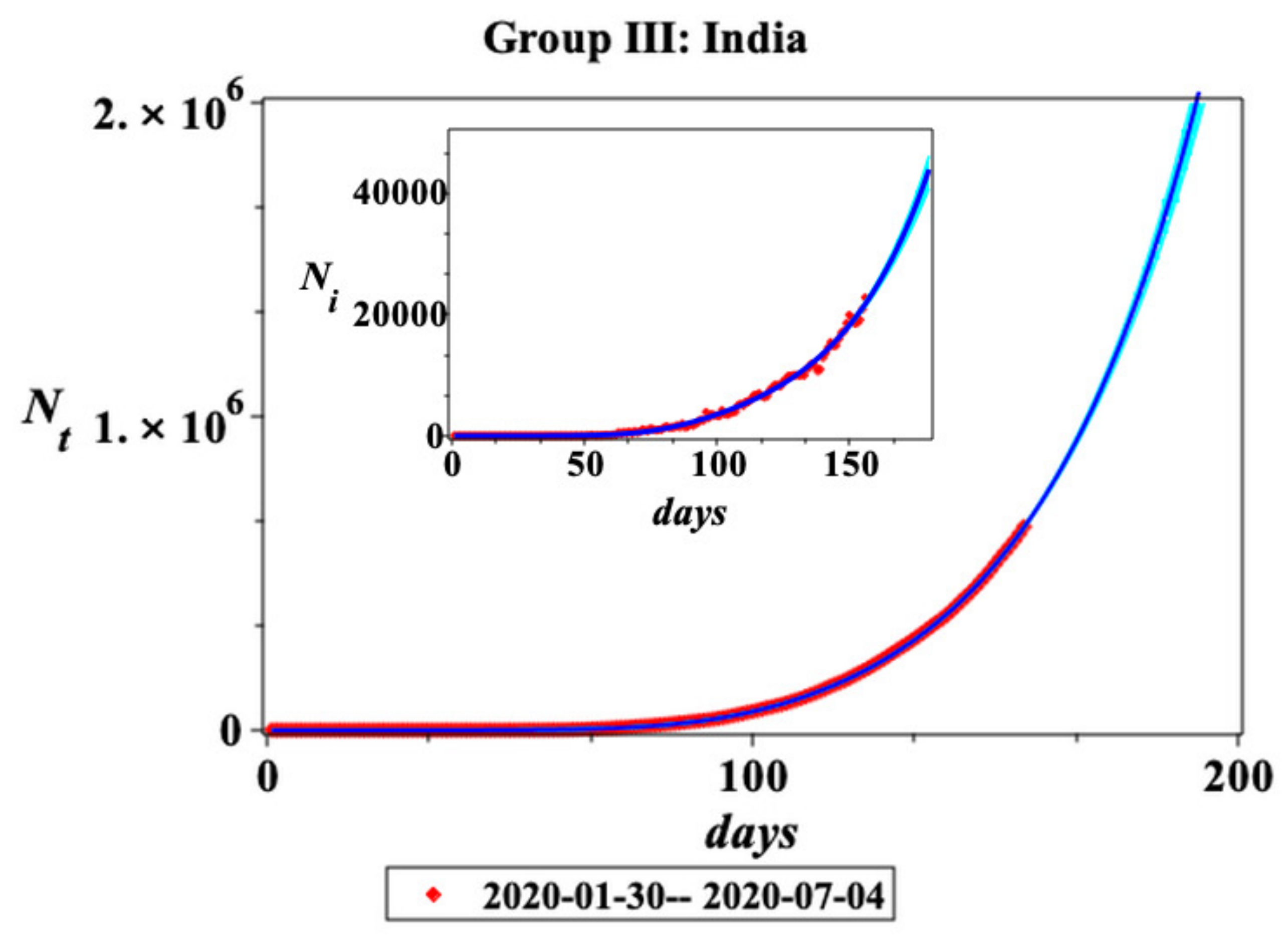}
         \end{subfigure}
         % \captionsetup{justification=RaggedRight}
    \caption{Theoretical outcomes of the configuration model: solid blue curves. Shadow cyan bands show the $95\%$ confidence interval. Actual data: red diamonds. Insets show the number of daily new cases.}
    \label{GroupI}
\end{figure*}

Comparison of Eqs. \eqref{Eq14a}  -- \eqref{Eq14b} with Eqs. \eqref{SIR1} -- \eqref{SIR3} of the SIR model shows that $N_t = I + R$,  $N_c$ should be identified with ${I S}/{N}$ and $\kappa$ with $ \nu$. After this identification, one can show that Eq. \eqref {Eq14a} can be written as a combination of Eqs. \eqref{SIR1} and \eqref{SIR2}. Thus, in our approach only one equation can be matched to the SIR model. Other equations, Eqs. \eqref{Eq14c} and  \eqref{Eq14b}, are new. 

The SIR model assumes that the population is thoroughly mixed, the individuals have the same number of contacts per day, and that transmission of the disease by contacts among individuals takes place with the same probability. None of these assumptions is realistic \cite{NMEJ2}.

Our model takes into account that the probability of an individual being infected may depend on the features of each pair of individuals, where one is infective and the other is susceptible, for instance, with a weak/strong immune system, and so forth. Besides, we omit the assumption that the population is mixed.

\section { Application to the COVID-19 pandemic}

To relate the time-dependent solution (\ref{Eq.9}) to the dynamics of the epidemic spreading, and extrapolate its future development, one should adjust the parameters $A$, $B$, $\alpha$, $\beta$ and $\mu$ with the data available for the known time trial domain. Besides, we have to specify the dependence of $\varepsilon(t)$ on time. To proceed, we take a simple but reasonable choice of a linear function, $\varepsilon(t) =\kappa t$. This choice provides a reasonable agreement of our results with most of the actual data. However, for the description of the epidemic spreading with a second epidemic wave, as has occurred for example with Iran, one can make a mre sophisticated choice of $\varepsilon  (t)$. We will show how to deal with this case below.

      \begin{figure}[tbh]
\centering
      \includegraphics[width=0.9\linewidth]{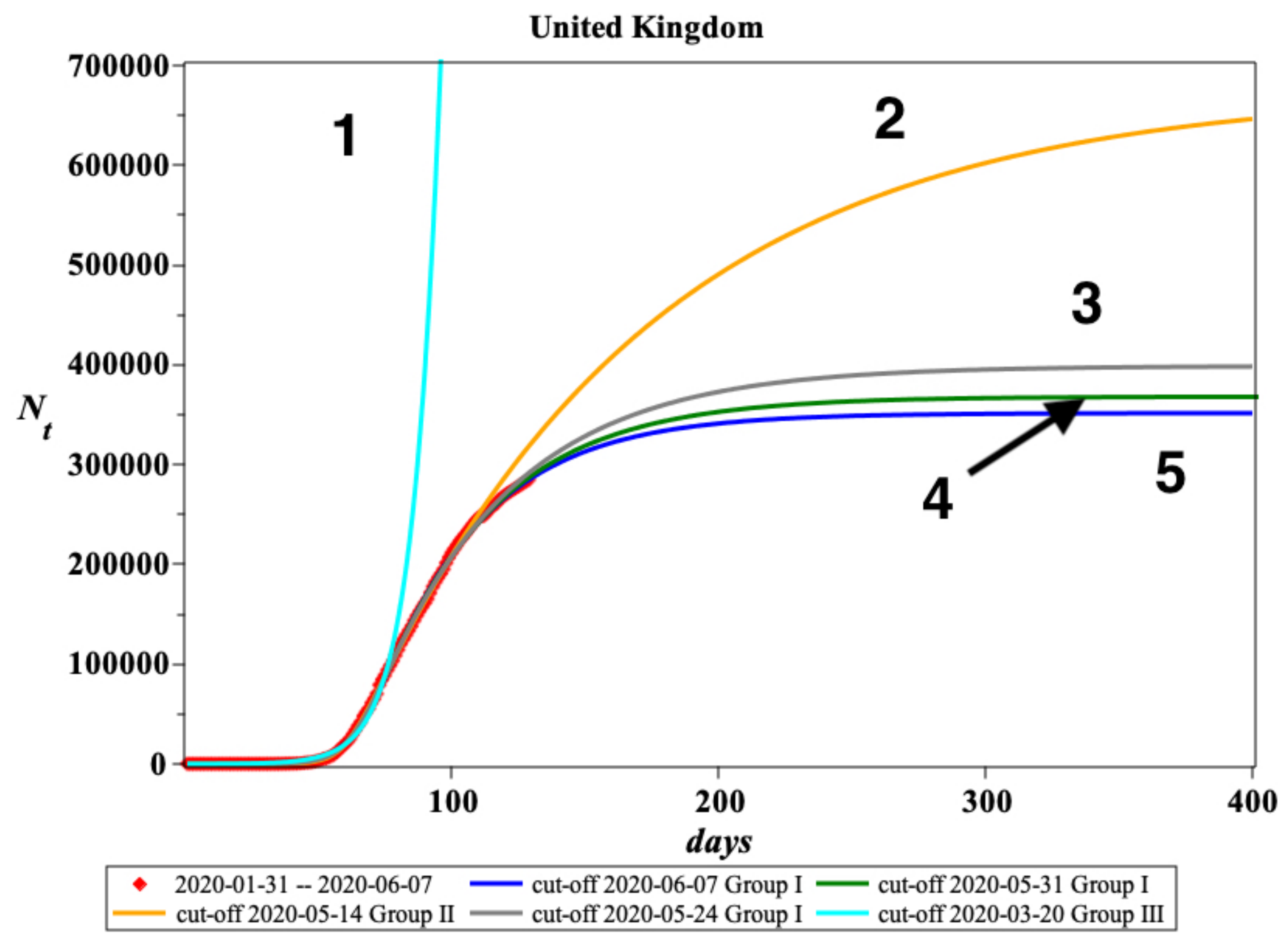}  
       %\captionsetup{justification=RaggedRight}
      	\caption{ Example of the country migration from one group to another during the epidemic spreading. Actual data: red diamonds. 	}
      \label{UK}
      \end{figure}
The COVID-19 data for this analysis are obtained from a publicly available database 
\cite{COR,OWD}. We made a comparison of our model with coronavirus data extracted for 213 
countries from the database \cite{OWD}. The detailed account of the comparison of our findings 
with the empirical data is presented in the Supplementary Material (SM).
For convenience, we define new variables: $\beta \varepsilon = \kappa t$ and $b = \beta \mu$. Then, Eq.\eqref{Eq.9} can be recast as
\begin{align}
N_t=  Ae^{\alpha (\kappa t -b)} \, \Phi(-e^{\kappa t-b},1,\alpha) +B.
\label{SM3a}
\end{align}
For each country, the constants $A$ and $B$ are obtained from the initial and final conditions.  
We understand the final condition as the total number of infected individuals at a cut-off date 
contained in the Coronavirus Pandemic (COVID-19) database \cite{OWD}.

We divide all countries into three groups. Group I: The countries where the epidemic is over or almost over. Group II: The countries where the epidemic is in progress, but available empirical data allow us to predict the epidemic end. Group III: The countries with a high-value of the critical parameter, $\alpha \geq1$, or with high-level fluctuations in the daily number of infected individuals. The insets show the number of daily new cases. 

In Fig. \ref{GroupI},  we compare the theoretical solutions of our model (solid blue curves) with the empirical data (red diamonds) for each of the groups. The insets show the number of daily new cases.  Shadow cyan bands show the $95 \%$ confidence interval. In Fig. \ref{UK}, we demonstrate the forecasting features of our model for the UK.  When the time trial domain increases (curves from 1 through 5), the unstable regime of our solution  ($\alpha>1$) changes to a stable regime ($\alpha<1$). Thus, during the forecasting period, the UK moves from the third group ($\alpha > 1$) to the first one ($\alpha < 0.9$).

	\begin{table}[tbh]   
\captionsetup{justification=raggedright}
        \caption{ The essential parameters of the configuration model}  
        
        \centering
\begin{tabular}{|l|c|c|c|c|c|c|c|}
  \hline
Country & Transmission rate &  Critical spreading \\
  & of epidemic $\kappa$  &  parameter  $\alpha$  \\
   \hline
Bangladesh & 0.072 & 1.026 \\
   \hline
Belgium &  0.214  & 0.787 \\
    \hline
Brazil &  0.075  & 1.108\\
  \hline
Canada & 0.135 & 0.787  \\
  \hline
Chile & 0.226  & 0.229 \\
  \hline
Colombia & 0.1 & 1.39 \\
    \hline
France & 0.371 &  0.9 \\
\hline
Germany & 0.335 &   0.901 \\
 \hline
India & 0.079 & 1.366 \\
 \hline
 Iran & 0.05 & 0.646 \\
 \hline
Italy & 0.207  &  0.8 \\
  \hline
Mexico & 0.072  & 1.15  \\
 \hline
Pakistan & 0.202 & 0.229 \\
  \hline
Peru & 0.094 & 0.746 \\
 \hline
Russia & 0.152 & 0.946 \\
 \hline
 South Africa & 0.04  & 2.094\\
 \hline
Spain & 0.25 &  0.8  \\
 \hline
Turkey & 0.386 &  0.946  \\
 \hline
UK &  0.157 & 0.827\\
 \hline
 USA &  0.883 & 1.004\\
 \hline
World & 0.285 & 1.04 \\
\hline
\end{tabular}
\end{table}

In Table 1, we present the essential parameters for the countries from the top 20: the 
transmission rate of epidemic $\kappa$ and the critical spreading parameter $\alpha$  with a 
cut-off date of 4 July 2020.

\begin{figure*}[t!]
   % \centering
    \begin{subfigure}[t]{0.32\textwidth}
        \centering
        \includegraphics[height=1.82in]{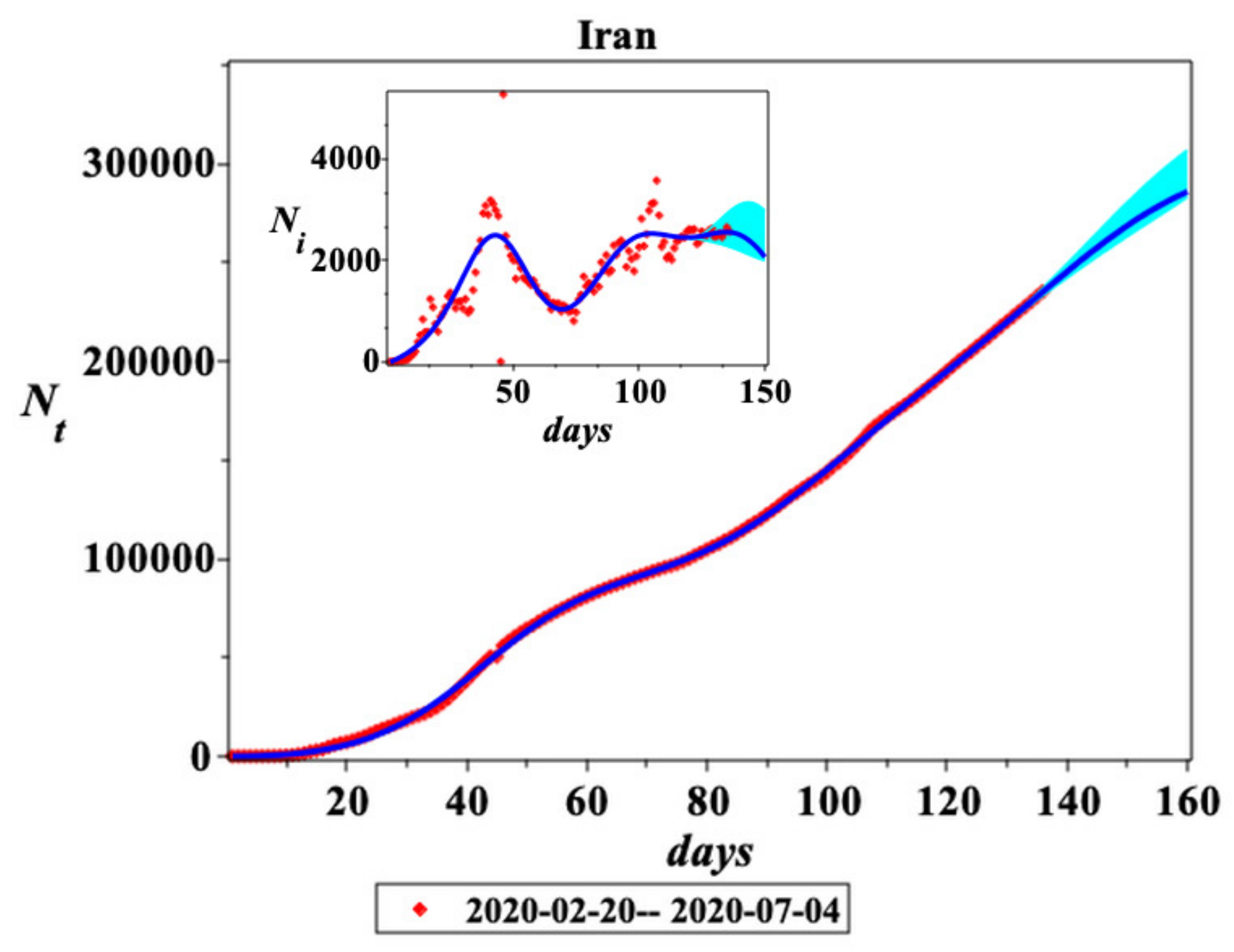}    
         \end{subfigure}
    \begin{subfigure}[t]{0.32\textwidth}
        \centering
        \includegraphics[height=1.9in]{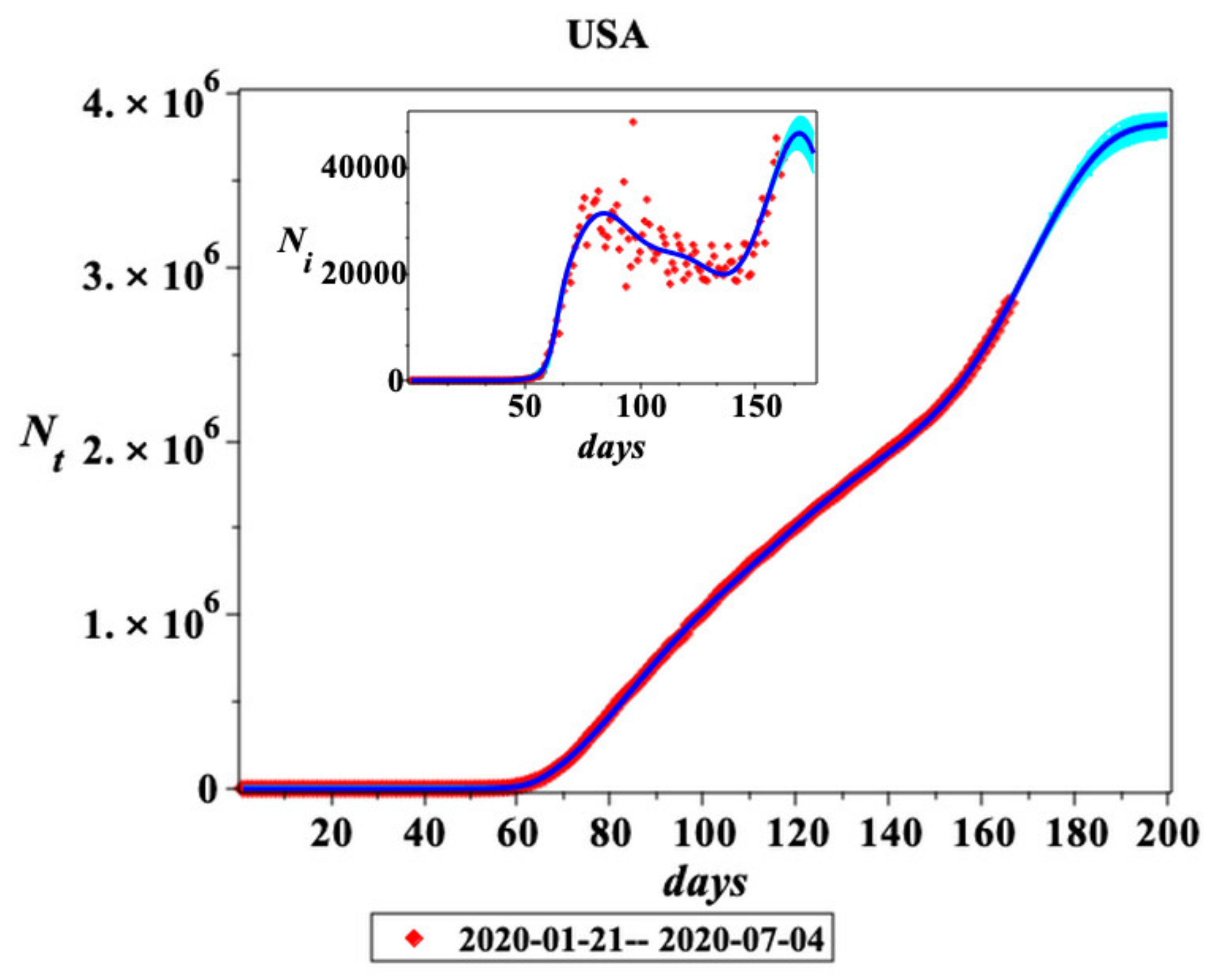}
         \end{subfigure}
          \begin{subfigure}[t]{0.32\textwidth}
        \centering
        \includegraphics[height=1.88in]{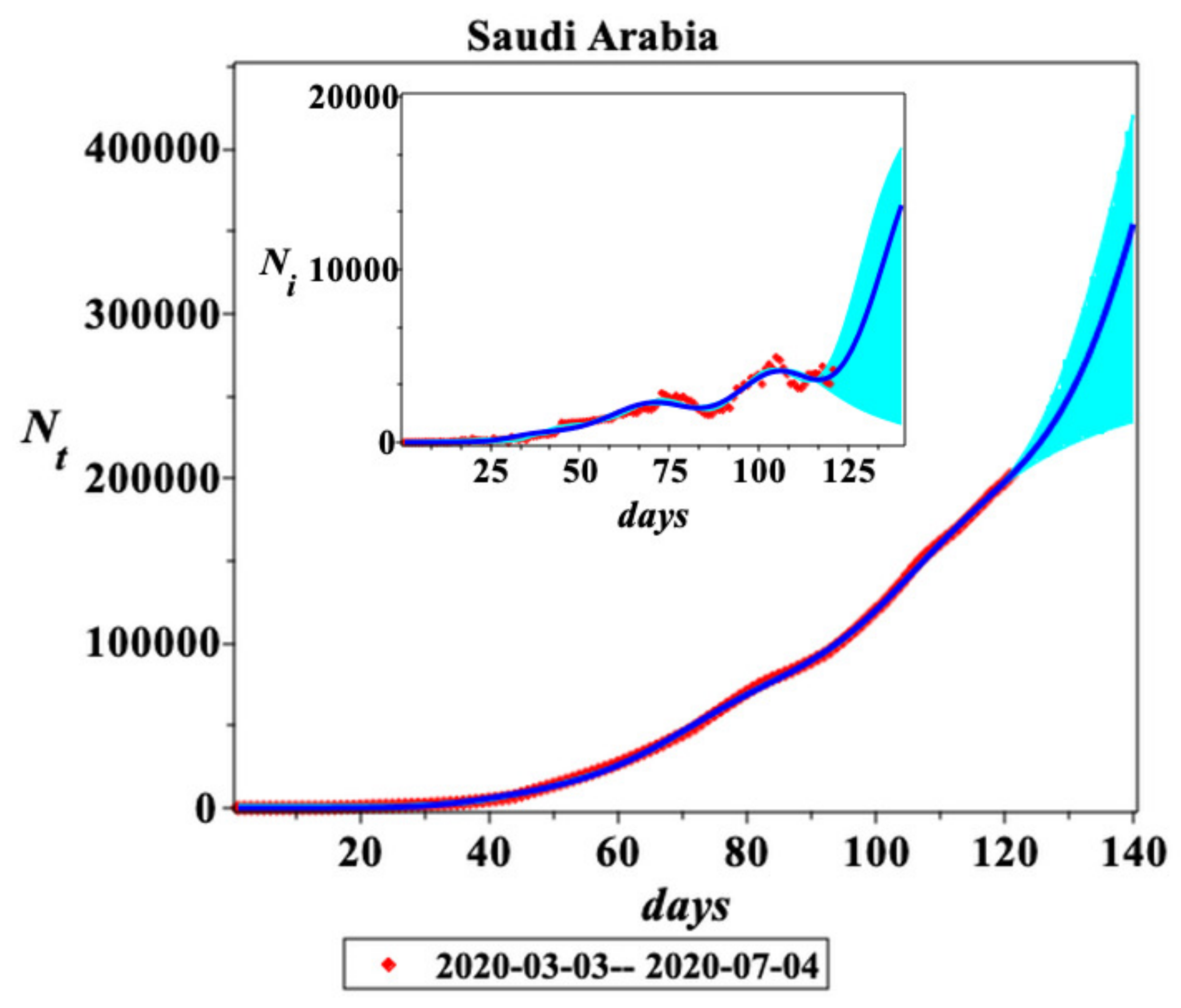}
         \end{subfigure}
         % \captionsetup{justification=RaggedRight}
    \caption{Second wave COVID-19.Theoretical outcomes of the configuration model: solid blue curves. Shadow cyan bands show the $95\%$ confidence interval. Actual data: red diamonds. Insets show the number of daily new cases.}
    \label{SW}
\end{figure*}

\subsection*{Second wave of COVID-19}

As mentioned above, the linear dependence of $\varepsilon$ on time can not describe the 
epidemic spreading with multiple waves. In order to describe this process, we  propose the 
following modification: $\varepsilon = \kappa t + \sum_n c_ne^{i\omega_n t}$, where $c_n$ 
and $\omega_n$ are additional fitting parameters. In Fig. \ref{SW}, we compare the theoretical 
outcomes of our configuration model with the empirical data for Iran, the United States and 
Saudi {Arabia}, with the choice of $\varepsilon (t)$ {as}
\begin{align}
	\varepsilon = \kappa t + \sum^2_{n =- 2} c_ne^{i\omega_n t}.
\end{align}
One can see that this choice of modulation of $\varepsilon(t)$ approximates the second wave effect well.

\section{ Conclusion and discussions} 

The accurate forecast procedures and predictive protocols for the COVID-19 pandemic spread 
require the knowledge of many factors, such as human behavior, reliable tests, and government 
regulations. In this situation, the importance of the mathematical models is in their ``predictive 
monitoring'' \cite{DDIL}. {This} implies that the predictions should be based on the dynamics of 
the epidemic spreading and changes in the real-world scenarios. 

The main problem of forecasting is the absence of reliable and accurate data on daily and total 
cases reported. It is important to stress that analytical solutions could be misleading for 
long-term forecasting because the parameters are going to change in the long run. Therefore 
the extrapolation and prediction should be taken with much care and, in most cases, only for a 
short period (see SM for details). 

The mathematical models that describe the epidemic spreading can be divided into two main 
groups: the stochastic and the deterministic models, written for time-dependent average 
variables. Our model is an intermediate one --- it is based on a set of differential equations 
obtained from the statistical physics of networks. We used the following assumptions in the 
derivation of our model: 1) the probability of an individual being infected through the 
disease-causing contact depends on the individual's features; 2) the epidemic spreads across 
the complex network in thermodynamical equilibrium. 

We have obtained analytical solutions which interpolate the daily and total numbers of infected individuals, and forecast the future epidemic development. These solutions describe both the stable epidemic development and the dynamics close to instability. This method, as applied to COVID-19, gives a reasonable picture for the pandemic development. Our approach can help to better understand the network's role in the epidemic spreading. Further progress in this direction will require to use of more advanced approaches, including those based on non-equilibrium statistical physics.
 
\acknowledgements

The work by G.P.B. was done at Los Alamos National Laboratory managed by Triad National Security, LLC, for the National Nuclear Security Administration of the U.S. Department of Energy under Contract No. 89233218CNA000001.


%\bibliography{nonass,GRG,DST,CN,COVID}

\begin{thebibliography}{34}%
\makeatletter
\providecommand \@ifxundefined [1]{%
 \@ifx{#1\undefined}
}%
\providecommand \@ifnum [1]{%
 \ifnum #1\expandafter \@firstoftwo
 \else \expandafter \@secondoftwo
 \fi
}%
\providecommand \@ifx [1]{%
 \ifx #1\expandafter \@firstoftwo
 \else \expandafter \@secondoftwo
 \fi
}%
\providecommand \natexlab [1]{#1}%
\providecommand \enquote  [1]{``#1''}%
\providecommand \bibnamefont  [1]{#1}%
\providecommand \bibfnamefont [1]{#1}%
\providecommand \citenamefont [1]{#1}%
\providecommand \href@noop [0]{\@secondoftwo}%
\providecommand \href [0]{\begingroup \@sanitize@url \@href}%
\providecommand \@href[1]{\@@startlink{#1}\@@href}%
\providecommand \@@href[1]{\endgroup#1\@@endlink}%
\providecommand \@sanitize@url [0]{\catcode `\\12\catcode `\$12\catcode
  `\&12\catcode `\#12\catcode `\^12\catcode `\_12\catcode `\%12\relax}%
\providecommand \@@startlink[1]{}%
\providecommand \@@endlink[0]{}%
\providecommand \url  [0]{\begingroup\@sanitize@url \@url }%
\providecommand \@url [1]{\endgroup\@href {#1}{\urlprefix }}%
\providecommand \urlprefix  [0]{URL }%
\providecommand \Eprint [0]{\href }%
\providecommand \doibase [0]{http://dx.doi.org/}%
\providecommand \selectlanguage [0]{\@gobble}%
\providecommand \bibinfo  [0]{\@secondoftwo}%
\providecommand \bibfield  [0]{\@secondoftwo}%
\providecommand \translation [1]{[#1]}%
\providecommand \BibitemOpen [0]{}%
\providecommand \bibitemStop [0]{}%
\providecommand \bibitemNoStop [0]{.\EOS\space}%
\providecommand \EOS [0]{\spacefactor3000\relax}%
\providecommand \BibitemShut  [1]{\csname bibitem#1\endcsname}%
\let\auto@bib@innerbib\@empty
%</preamble>
\bibitem [{\citenamefont {Adam}(2020)}]{AD}%
  \BibitemOpen
  \bibfield  {author} {\bibinfo {author} {\bibfnamefont {David}\ \bibnamefont
  {Adam}},\ }\bibfield  {title} {\enquote {\bibinfo {title} {{Special report:
  The simulations driving the world's response to COVID-19.}}}\ }\href@noop {}
  {\bibfield  {journal} {\bibinfo  {journal} {Nature}\ }\textbf {\bibinfo
  {volume} {580}},\ \bibinfo {pages} {316--318} (\bibinfo {year}
  {2020})}\BibitemShut {NoStop}%
\bibitem [{\citenamefont {Ivorra}\ \emph {et~al.}(2020)\citenamefont {Ivorra},
  \citenamefont {Ferr\'andez}, \citenamefont {Vela-P\'erez},\ and\
  \citenamefont {Ramos}}]{IBMRF}%
  \BibitemOpen
  \bibfield  {author} {\bibinfo {author} {\bibfnamefont {B.}~\bibnamefont
  {Ivorra}}, \bibinfo {author} {\bibfnamefont {M.R.}\ \bibnamefont
  {Ferr\'andez}}, \bibinfo {author} {\bibfnamefont {M.}~\bibnamefont
  {Vela-P\'erez}}, \ and\ \bibinfo {author} {\bibfnamefont {A.M.}\ \bibnamefont
  {Ramos}},\ }\bibfield  {title} {\enquote {\bibinfo {title} {{Mathematical
  modeling of the spread of the coronavirus disease 2019 (COVID-19) taking into
  account the undetected infections. The case of China}},}\ }\href {\doibase
  https://doi.org/10.1016/j.cnsns.2020.105303} {\bibfield  {journal} {\bibinfo
  {journal} {{Communications in Nonlinear Science and Numerical Simulation}}\
  }\textbf {\bibinfo {volume} {88}},\ \bibinfo {pages} {105303} (\bibinfo
  {year} {2020})}\BibitemShut {NoStop}%
\bibitem [{\citenamefont {Cobey}(2020)}]{CS}%
  \BibitemOpen
  \bibfield  {author} {\bibinfo {author} {\bibfnamefont {Sarah}\ \bibnamefont
  {Cobey}},\ }\bibfield  {title} {\enquote {\bibinfo {title} {Modeling
  infectious disease dynamics},}\ }\href {\doibase 10.1126/science.abb5659}
  {\bibfield  {journal} {\bibinfo  {journal} {Science}\ }\textbf {\bibinfo
  {volume} {368}},\ \bibinfo {pages} {713--714} (\bibinfo {year}
  {2020})}\BibitemShut {NoStop}%
\bibitem [{\citenamefont {Bollob\'as}(2001)}]{BB1}%
  \BibitemOpen
  \bibfield  {author} {\bibinfo {author} {\bibfnamefont {B\'ela}\ \bibnamefont
  {Bollob\'as}},\ }\href@noop {} {\emph {\bibinfo {title} {Random Graphs}}},\
  \bibinfo {edition} {2nd}\ ed.,\ Cambridge studies in advanced mathematics 73\
  (\bibinfo  {publisher} {Cambridge University Press},\ \bibinfo {year}
  {2001})\BibitemShut {NoStop}%
\bibitem [{\citenamefont {Newman}(2002)}]{NMEJ2}%
  \BibitemOpen
  \bibfield  {author} {\bibinfo {author} {\bibfnamefont {M.~E.~J.}\
  \bibnamefont {Newman}},\ }\bibfield  {title} {\enquote {\bibinfo {title}
  {Spread of epidemic disease on networks},}\ }\href {\doibase
  10.1103/PhysRevE.66.016128} {\bibfield  {journal} {\bibinfo  {journal} {Phys.
  Rev. E}\ }\textbf {\bibinfo {volume} {66}},\ \bibinfo {pages} {016128}
  (\bibinfo {year} {2002})}\BibitemShut {NoStop}%
\bibitem [{\citenamefont {Moore}\ and\ \citenamefont {Rogers}(2020)}]{MSRT}%
  \BibitemOpen
  \bibfield  {author} {\bibinfo {author} {\bibfnamefont {Sam}\ \bibnamefont
  {Moore}}\ and\ \bibinfo {author} {\bibfnamefont {Tim}\ \bibnamefont
  {Rogers}},\ }\bibfield  {title} {\enquote {\bibinfo {title} {{Predicting the
  Speed of Epidemics Spreading in Networks}},}\ }\href {\doibase
  10.1103/PhysRevLett.124.068301} {\bibfield  {journal} {\bibinfo  {journal}
  {Phys. Rev. Lett.}\ }\textbf {\bibinfo {volume} {124}},\ \bibinfo {pages}
  {068301} (\bibinfo {year} {2020})}\BibitemShut {NoStop}%
\bibitem [{\citenamefont {Lloyd}\ and\ \citenamefont {May}(2001)}]{LAMR}%
  \BibitemOpen
  \bibfield  {author} {\bibinfo {author} {\bibfnamefont {Alun~L.}\ \bibnamefont
  {Lloyd}}\ and\ \bibinfo {author} {\bibfnamefont {Robert~M.}\ \bibnamefont
  {May}},\ }\bibfield  {title} {\enquote {\bibinfo {title} {{How Viruses Spread
  among Computers and People}},}\ }\href {http://www.jstor.org/stable/3083757}
  {\bibfield  {journal} {\bibinfo  {journal} {Science}\ }\textbf {\bibinfo
  {volume} {292}},\ \bibinfo {pages} {1316--1317} (\bibinfo {year}
  {2001})}\BibitemShut {NoStop}%
\bibitem [{\citenamefont {Barabasi}(2016)}]{BAL}%
  \BibitemOpen
  \bibfield  {author} {\bibinfo {author} {\bibfnamefont {Albert-Laszlo}\
  \bibnamefont {Barabasi}},\ }\href
  {http://gen.lib.rus.ec/book/index.php?md5=625AFD86B4E4550057469ECBF69826C8}
  {\emph {\bibinfo {title} {Network Science}}}\ (\bibinfo  {publisher}
  {Cambridge University Press},\ \bibinfo {year} {2016})\BibitemShut {NoStop}%
\bibitem [{\citenamefont {{S. N. Dorogovtsev and J. F. F. Mendes
  }}(2003)}]{DSMF}%
  \BibitemOpen
  \bibfield  {author} {\bibinfo {author} {\bibnamefont {{S. N. Dorogovtsev and
  J. F. F. Mendes }}},\ }\href@noop {} {\emph {\bibinfo {title} {{Evolution of
  Networks: From Biological Nets to the Internet and WWW}}}}\ (\bibinfo
  {publisher} {Oxford University Press},\ \bibinfo {year} {2003})\BibitemShut
  {NoStop}%
\bibitem [{\citenamefont {Barrat}\ \emph {et~al.}(2008)\citenamefont {Barrat},
  \citenamefont {Barthelemy},\ and\ \citenamefont {Vespignani}}]{BABM}%
  \BibitemOpen
  \bibfield  {author} {\bibinfo {author} {\bibfnamefont {A.}~\bibnamefont
  {Barrat}}, \bibinfo {author} {\bibfnamefont {M.}~\bibnamefont {Barthelemy}},
  \ and\ \bibinfo {author} {\bibfnamefont {A.}~\bibnamefont {Vespignani}},\
  }\href@noop {} {\emph {\bibinfo {title} {{Dynamical Processes on Complex
  Networks}}}}\ (\bibinfo  {publisher} {{Cambridge University Press}},\
  \bibinfo {year} {2008})\BibitemShut {NoStop}%
\bibitem [{\citenamefont {Newman}(2018)}]{MN2018}%
  \BibitemOpen
  \bibfield  {author} {\bibinfo {author} {\bibfnamefont {Mark}\ \bibnamefont
  {Newman}},\ }\href@noop {} {\emph {\bibinfo {title} {Networks}}}\ (\bibinfo
  {publisher} {Oxford University Press},\ \bibinfo {year} {2018})\BibitemShut
  {NoStop}%
\bibitem [{\citenamefont {Caldarelli}(2007)}]{GCal}%
  \BibitemOpen
  \bibfield  {author} {\bibinfo {author} {\bibfnamefont {Guido}\ \bibnamefont
  {Caldarelli}},\ }\href@noop {} {\emph {\bibinfo {title} {Scale-free networks:
  complex webs in nature and technology}}}\ (\bibinfo  {publisher} {Oxford
  University Press, USA},\ \bibinfo {year} {2007})\BibitemShut {NoStop}%
 \bibitem [{\citenamefont {May}\ and\ \citenamefont {Lloyd}(2001)}]{MRLA}%
   \BibitemOpen
   \bibfield  {author} {\bibinfo {author} {\bibfnamefont {Robert~M.}\
   \bibnamefont {May}}\ and\ \bibinfo {author} {\bibfnamefont {Alun~L.}\
   \bibnamefont {Lloyd}},\ }\bibfield  {title}  {\enquote {\bibinfo {title} {Infection dynamics on 
   scale-free networks},}\ } \href {\doibase 10.1103/PhysRevE.64.066112}   {\bibfield  {journal} 
   {\bibinfo
   {journal} {Phys. Rev. E}\ }\textbf {\bibinfo {volume} {64}},\ \bibinfo {pages}
   {066112} (\bibinfo {year} {2001})}\BibitemShut {NoStop}%
\bibitem [{\citenamefont {Barab{\'a}si}(2009)}]{BAL1}%
  \BibitemOpen
  \bibfield  {author} {\bibinfo {author} {\bibfnamefont
  {Albert-L{\'a}szl{\'o}}\ \bibnamefont {Barab{\'a}si}},\ }\bibfield  {title}
  {\enquote {\bibinfo {title} {{Scale-Free Networks: A Decade and Beyond}},}\
  }\href {\doibase 10.1126/science.1173299} {\bibfield  {journal} {\bibinfo
  {journal} {Science}\ }\textbf {\bibinfo {volume} {325}},\ \bibinfo {pages}
  {412--413} (\bibinfo {year} {2009})}\BibitemShut {NoStop}%
\bibitem [{\citenamefont {Girvan}\ and\ \citenamefont {Newman}(2002)}]{GMNM}%
  \BibitemOpen
  \bibfield  {author} {\bibinfo {author} {\bibfnamefont {M.}~\bibnamefont
  {Girvan}}\ and\ \bibinfo {author} {\bibfnamefont {M.~E.~J.}\ \bibnamefont
  {Newman}},\ }\bibfield  {title} {\enquote {\bibinfo {title} {Community
  structure in social and biological networks},}\ }\href {\doibase
  10.1073/pnas.122653799} {\bibfield  {journal} {\bibinfo  {journal}
  {Proceedings of the National Academy of Sciences}\ }\textbf {\bibinfo
  {volume} {99}},\ \bibinfo {pages} {7821--7826} (\bibinfo {year}
  {2002})}\BibitemShut {NoStop}%
\bibitem [{\citenamefont {Albert}\ and\ \citenamefont
  {Barab\'asi}(2002)}]{ARB}%
  \BibitemOpen
  \bibfield  {author} {\bibinfo {author} {\bibfnamefont {R\'eka}\ \bibnamefont
  {Albert}}\ and\ \bibinfo {author} {\bibfnamefont {Albert-L\'aszl\'o}\
  \bibnamefont {Barab\'asi}},\ }\bibfield  {title} {\enquote {\bibinfo {title}
  {Statistical mechanics of complex networks},}\ }\href@noop {} {\bibfield
  {journal} {\bibinfo  {journal} {Rev. Mod. Phys.}\ }\textbf {\bibinfo {volume}
  {74}},\ \bibinfo {pages} {47--97} (\bibinfo {year} {2002})}\BibitemShut
  {NoStop}%
\bibitem [{\citenamefont {Newman}(2003)}]{MEJN1}%
  \BibitemOpen
  \bibfield  {author} {\bibinfo {author} {\bibfnamefont {M.~E.~J.}\
  \bibnamefont {Newman}},\ }\bibfield  {title} {\enquote {\bibinfo {title} {The
  structure and function of complex networks},}\ }\href@noop {} {\bibfield
  {journal} {\bibinfo  {journal} {SIAM Review}\ }\textbf {\bibinfo {volume}
  {45}},\ \bibinfo {pages} {167--256} (\bibinfo {year} {2003})}\BibitemShut
  {NoStop}%
\bibitem [{\citenamefont {Romualdo Pastor-Satorras}(2003)}]{RPMR}%
  \BibitemOpen
  \bibfield  {author} {\bibinfo {author} {\bibfnamefont {Albert
  Diaz-Guilera~(eds.)}\ \bibnamefont {Romualdo Pastor-Satorras}, \bibfnamefont
  {Miguel~Rubi}},\ }\href@noop {} {\emph {\bibinfo {title} {{Statistical
  Mechanics of Complex Networks}}}},\ Lecture Notes in Physics 625\ (\bibinfo
  {publisher} {Springer-Verlag Berlin Heidelberg},\ \bibinfo {year}
  {2003})\BibitemShut {NoStop}%
\bibitem [{\citenamefont {Park}\ and\ \citenamefont {Newman}(2004)}]{PJNM}%
  \BibitemOpen
  \bibfield  {author} {\bibinfo {author} {\bibfnamefont {Juyong}\ \bibnamefont
  {Park}}\ and\ \bibinfo {author} {\bibfnamefont {M.~E.~J.}\ \bibnamefont
  {Newman}},\ }\bibfield  {title} {\enquote {\bibinfo {title} {Statistical
  mechanics of networks},}\ }\href@noop {} {\bibfield  {journal} {\bibinfo
  {journal} {Phys. Rev. E}\ }\textbf {\bibinfo {volume} {70}},\ \bibinfo
  {pages} {066117} (\bibinfo {year} {2004})}\BibitemShut {NoStop}%
\bibitem [{\citenamefont {Garlaschelli}\ and\ \citenamefont
  {Loffredo}(2006)}]{CDLM}%
  \BibitemOpen
  \bibfield  {author} {\bibinfo {author} {\bibfnamefont {Diego}\ \bibnamefont
  {Garlaschelli}}\ and\ \bibinfo {author} {\bibfnamefont {Maria~I.}\
  \bibnamefont {Loffredo}},\ }\bibfield  {title} {\enquote {\bibinfo {title}
  {Multispecies grand-canonical models for networks with reciprocity},}\ }\href
  {\doibase 10.1103/PhysRevE.73.015101} {\bibfield  {journal} {\bibinfo
  {journal} {Phys. Rev. E}\ }\textbf {\bibinfo {volume} {73}},\ \bibinfo
  {pages} {015101} (\bibinfo {year} {2006})}\BibitemShut {NoStop}%
\bibitem [{\citenamefont {Cimini}\ \emph {et~al.}(2019)\citenamefont {Cimini},
  \citenamefont {Squartini}, \citenamefont {Saracco}, \citenamefont
  {Garlaschelli}, \citenamefont {Gabrielli},\ and\ \citenamefont
  {Caldarelli}}]{CGST}%
  \BibitemOpen
  \bibfield  {author} {\bibinfo {author} {\bibfnamefont {Giulio}\ \bibnamefont
  {Cimini}}, \bibinfo {author} {\bibfnamefont {Tiziano}\ \bibnamefont
  {Squartini}}, \bibinfo {author} {\bibfnamefont {Fabio}\ \bibnamefont
  {Saracco}}, \bibinfo {author} {\bibfnamefont {Diego}\ \bibnamefont
  {Garlaschelli}}, \bibinfo {author} {\bibfnamefont {Andrea}\ \bibnamefont
  {Gabrielli}}, \ and\ \bibinfo {author} {\bibfnamefont {Guido}\ \bibnamefont
  {Caldarelli}},\ }\bibfield  {title} {\enquote {\bibinfo {title} {The
  statistical physics of real-world networks},}\ }\href {\doibase
  10.1038/s42254-018-0002-6} {\bibfield  {journal} {\bibinfo  {journal} {Nature
  Reviews Physics}\ }\textbf {\bibinfo {volume} {1}},\ \bibinfo {pages}
  {58--71} (\bibinfo {year} {2019})}\BibitemShut {NoStop}%
\bibitem [{\citenamefont {Watts}\ and\ \citenamefont {Strogatz}(1998)}]{WDSS}%
  \BibitemOpen
  \bibfield  {author} {\bibinfo {author} {\bibfnamefont {Duncan~J.}\
  \bibnamefont {Watts}}\ and\ \bibinfo {author} {\bibfnamefont {Steven~H.}\
  \bibnamefont {Strogatz}},\ }\bibfield  {title} {\enquote {\bibinfo {title}
  {Collective dynamics of `small-world' networks},}\ }\href@noop {} {\bibfield
  {journal} {\bibinfo  {journal} {Nature}\ }\textbf {\bibinfo {volume} {393}},\
  \bibinfo {pages} {440 -- 442} (\bibinfo {year} {1998})}\BibitemShut {NoStop}%
\bibitem [{\citenamefont {Newman}\ \emph {et~al.}(2001)\citenamefont {Newman},
  \citenamefont {Strogatz},\ and\ \citenamefont {Watts}}]{NMSW}%
  \BibitemOpen
  \bibfield  {author} {\bibinfo {author} {\bibfnamefont {M.~E.~J.}\
  \bibnamefont {Newman}}, \bibinfo {author} {\bibfnamefont {S.~H.}\
  \bibnamefont {Strogatz}}, \ and\ \bibinfo {author} {\bibfnamefont {D.~J.}\
  \bibnamefont {Watts}},\ }\bibfield  {title} {\enquote {\bibinfo {title}
  {Random graphs with arbitrary degree distributions and their applications},}\
  }\href@noop {} {\bibfield  {journal} {\bibinfo  {journal} {Phys. Rev. E}\
  }\textbf {\bibinfo {volume} {64}},\ \bibinfo {pages} {026118} (\bibinfo
  {year} {2001})}\BibitemShut {NoStop}%
\bibitem [{\citenamefont {Boccaletti}\ \emph {et~al.}(2006)\citenamefont
  {Boccaletti}, \citenamefont {Latora}, \citenamefont {Moreno}, \citenamefont
  {Chavez},\ and\ \citenamefont {Hwang}}]{BSLV}%
  \BibitemOpen
  \bibfield  {author} {\bibinfo {author} {\bibfnamefont {S.}~\bibnamefont
  {Boccaletti}}, \bibinfo {author} {\bibfnamefont {V.}~\bibnamefont {Latora}},
  \bibinfo {author} {\bibfnamefont {Y.}~\bibnamefont {Moreno}}, \bibinfo
  {author} {\bibfnamefont {M.}~\bibnamefont {Chavez}}, \ and\ \bibinfo {author}
  {\bibfnamefont {D.-U.}\ \bibnamefont {Hwang}},\ }\bibfield  {title} {\enquote
  {\bibinfo {title} {Complex networks: Structure and dynamics},}\ }\href
  {\doibase https://doi.org/10.1016/j.physrep.2005.10.009} {\bibfield
  {journal} {\bibinfo  {journal} {Physics Reports}\ }\textbf {\bibinfo {volume}
  {424}},\ \bibinfo {pages} {175 -- 308} (\bibinfo {year} {2006})}\BibitemShut
  {NoStop}%
\bibitem [{\citenamefont {Garlaschelli}\ \emph {et~al.}(2013)\citenamefont
  {Garlaschelli}, \citenamefont {Ahnert}, \citenamefont {Fink},\ and\
  \citenamefont {Caldarelli}}]{CDAS}%
  \BibitemOpen
  \bibfield  {author} {\bibinfo {author} {\bibfnamefont {Diego}\ \bibnamefont
  {Garlaschelli}}, \bibinfo {author} {\bibfnamefont {Sebastian~E.}\
  \bibnamefont {Ahnert}}, \bibinfo {author} {\bibfnamefont {Thomas M.~A.}\
  \bibnamefont {Fink}}, \ and\ \bibinfo {author} {\bibfnamefont {Guido}\
  \bibnamefont {Caldarelli}},\ }\bibfield  {title} {\enquote {\bibinfo {title}
  {Low-temperature behaviour of social and economic networks},}\ }\href
  {\doibase 10.3390/e15083238} {\bibfield  {journal} {\bibinfo  {journal}
  {Entropy}\ }\textbf {\bibinfo {volume} {15}},\ \bibinfo {pages} {3148--3169}
  (\bibinfo {year} {2013})}\BibitemShut {NoStop}%
\bibitem [{\citenamefont {Wang}\ \emph {et~al.}(2017)\citenamefont {Wang},
  \citenamefont {Wilson},\ and\ \citenamefont {Hancock}}]{WJWR}%
  \BibitemOpen
  \bibfield  {author} {\bibinfo {author} {\bibfnamefont {Jianjia}\ \bibnamefont
  {Wang}}, \bibinfo {author} {\bibfnamefont {Richard~C}\ \bibnamefont
  {Wilson}}, \ and\ \bibinfo {author} {\bibfnamefont {Edwin~R}\ \bibnamefont
  {Hancock}},\ }\bibfield  {title} {\enquote {\bibinfo {title} {{Spin
  statistics, partition functions and network entropy}},}\ }\href {\doibase
  10.1093/comnet/cnx017} {\bibfield  {journal} {\bibinfo  {journal} {Journal of
  Complex Networks}\ }\textbf {\bibinfo {volume} {5}},\ \bibinfo {pages}
  {858--883} (\bibinfo {year} {2017})}\BibitemShut {NoStop}%
\bibitem [{\citenamefont {Fred~Brauer}(2012)}]{FBCC}%
  \BibitemOpen
  \bibfield  {author} {\bibinfo {author} {\bibfnamefont {Carlos
  Castillo-Chavez~(auth.)}\ \bibnamefont {Fred~Brauer}},\ }\href@noop {} {\emph
  {\bibinfo {title} {Mathematical models in population biology and
  epidemiology}}}\ (\bibinfo  {publisher} {Springer-Verlag New York},\ \bibinfo
  {year} {2012})\BibitemShut {NoStop}%
\bibitem [{\citenamefont {{Frank W. J. Olver, Daniel W. Lozier, Ronald F.
  Boisvert, Charles W. Clark}}(2010)}]{NIST}%
  \BibitemOpen
  \bibfield  {author} {\bibinfo {author} {\bibnamefont {{Frank W. J. Olver,
  Daniel W. Lozier, Ronald F. Boisvert, Charles W. Clark}}},\ }\href@noop {}
  {\emph {\bibinfo {title} {{NIST Handbook of Mathematical Functions}}}}\
  (\bibinfo  {publisher} {Cambridge University Press},\ \bibinfo {year}
  {2010})\BibitemShut {NoStop}%
\bibitem [{\citenamefont {Martcheva}(2015)}]{MM1}%
  \BibitemOpen
  \bibfield  {author} {\bibinfo {author} {\bibfnamefont {Maia}\ \bibnamefont
  {Martcheva}},\ }\href@noop {} {\emph {\bibinfo {title} {{An Introduction to
  Mathematical Epidemiology}}}},\ Texts in Applied Mathematics 61\ (\bibinfo
  {publisher} {Springer US},\ \bibinfo {year} {2015})\BibitemShut {NoStop}%
\bibitem [{\citenamefont {Meyers}(2009)}]{MRA}%
  \BibitemOpen
  \bibfield  {author} {\bibinfo {author} {\bibfnamefont {Robert~A.}\
  \bibnamefont {Meyers}},\ }\href@noop {} {\emph {\bibinfo {title}
  {{Encyclopedia of Complexity and Systems Science}}}}\ (\bibinfo  {publisher}
  {Springer},\ \bibinfo {year} {2009})\BibitemShut {NoStop}%
\bibitem [{\citenamefont {Brauer}(2019)}]{BFR}%
  \BibitemOpen
  \bibfield  {author} {\bibinfo {author} {\bibfnamefont {Fred}\ \bibnamefont
  {Brauer}},\ }\href@noop {} {\emph {\bibinfo {title} {{Mathematical Models in
  Epidemiology}}}}\ (\bibinfo  {publisher} {Springer},\ \bibinfo {year}
  {2019})\BibitemShut {NoStop}%
\bibitem [{\citenamefont {https://www.worldometers.info/coronavirus}()}]{COR}%
  \BibitemOpen
  \bibfield  {author} {\bibinfo {author} {\bibnamefont
  {https://www.worldometers.info/coronavirus}},\ }\href@noop {} {\bibinfo
  {journal} {{COVID-19 Coronavirus Pandemic}}\ }\BibitemShut {NoStop}%
\bibitem [{\citenamefont {Roser}\ \emph {et~al.}(2020)\citenamefont {Roser},
  \citenamefont {Ritchie}, \citenamefont {Ortiz-Ospina},\ and\ \citenamefont
  {Hasell}}]{OWD}%
  \BibitemOpen
\bibfield  {journal} {  }\bibfield  {author} {\bibinfo {author} {\bibfnamefont
  {Max}\ \bibnamefont {Roser}}, \bibinfo {author} {\bibfnamefont {Hannah}\
  \bibnamefont {Ritchie}}, \bibinfo {author} {\bibfnamefont {Esteban}\
  \bibnamefont {Ortiz-Ospina}}, \ and\ \bibinfo {author} {\bibfnamefont {Joe}\
  \bibnamefont {Hasell}},\ }\bibfield  {title} {\enquote {\bibinfo {title}
  {{Coronavirus Pandemic (COVID-19) }},}\ }\href@noop {} {\bibfield  {journal}
  {\bibinfo  {journal} {Our World in Data}\ } (\bibinfo {year} {2020})},\
  \bibinfo {note} {https://ourworldindata.org/coronavirus}\BibitemShut
  {NoStop}%
\bibitem [{\citenamefont {{Jianxi Luo}}(2020)}]{DDIL}%
  \BibitemOpen
  \bibfield  {author} {\bibinfo {author} {\bibnamefont {{Jianxi Luo}}},\
  }\bibfield  {title} {\enquote {\bibinfo {title} {{Predictive Monitoring of
  COVID-19}},}\ }\href@noop {} {\bibfield  {journal} {\bibinfo  {journal}
  {{Data-Driven Innovation Lab: https://people.sutd.edu.sg/}}\ } (\bibinfo
  {year} {2020})}\BibitemShut {NoStop}%
\end{thebibliography}

\appendix

\section*{Supplemental Material}

 In this Supplemental Material we explain the calculations, approximations, and intermediate  steps of the main text. The total number of infected individuals in our model is described by the the Lerch 
 Transcendent:
 \begin{align}
 N_t  = A e^{\alpha \beta (\varepsilon (t) - \mu)} \Phi\big (- e^{\beta (\varepsilon (t)  
 -\mu)} ,1, \alpha\big ) + B,
 \label{SM1}
 \end{align}
  and the daily number of infected individuals, $N_i$, is obtained by taking the derivative of 
 $N_t$,
 \begin{align}
 N_i =\frac{dN_t}{d t} .
 \label{SM2}
 \end{align}
 When the epidemic does not generate multiple waves, a linear dependence of $\varepsilon$ on 
 time yields good results. It is convenient to define new variables: $\beta 
 \varepsilon = \kappa t$ and $b = \beta \mu$. (Note that in  the SM we consider only the  
 linear dependence of $\varepsilon$ on $t$.) Then, Eq.\eqref{SM1} can be recast as
 \begin{align}
 N_t=  Ae^{\alpha (\kappa t -b)} \, \Phi(-e^{\kappa t-b},1,\alpha) +B.
 \label{SM3}
 \end{align}
  
\begin{figure*}[t!]
   % \centering
    \begin{subfigure}[t]{0.8\textwidth}
        \centering
        \includegraphics[height=3.8in]{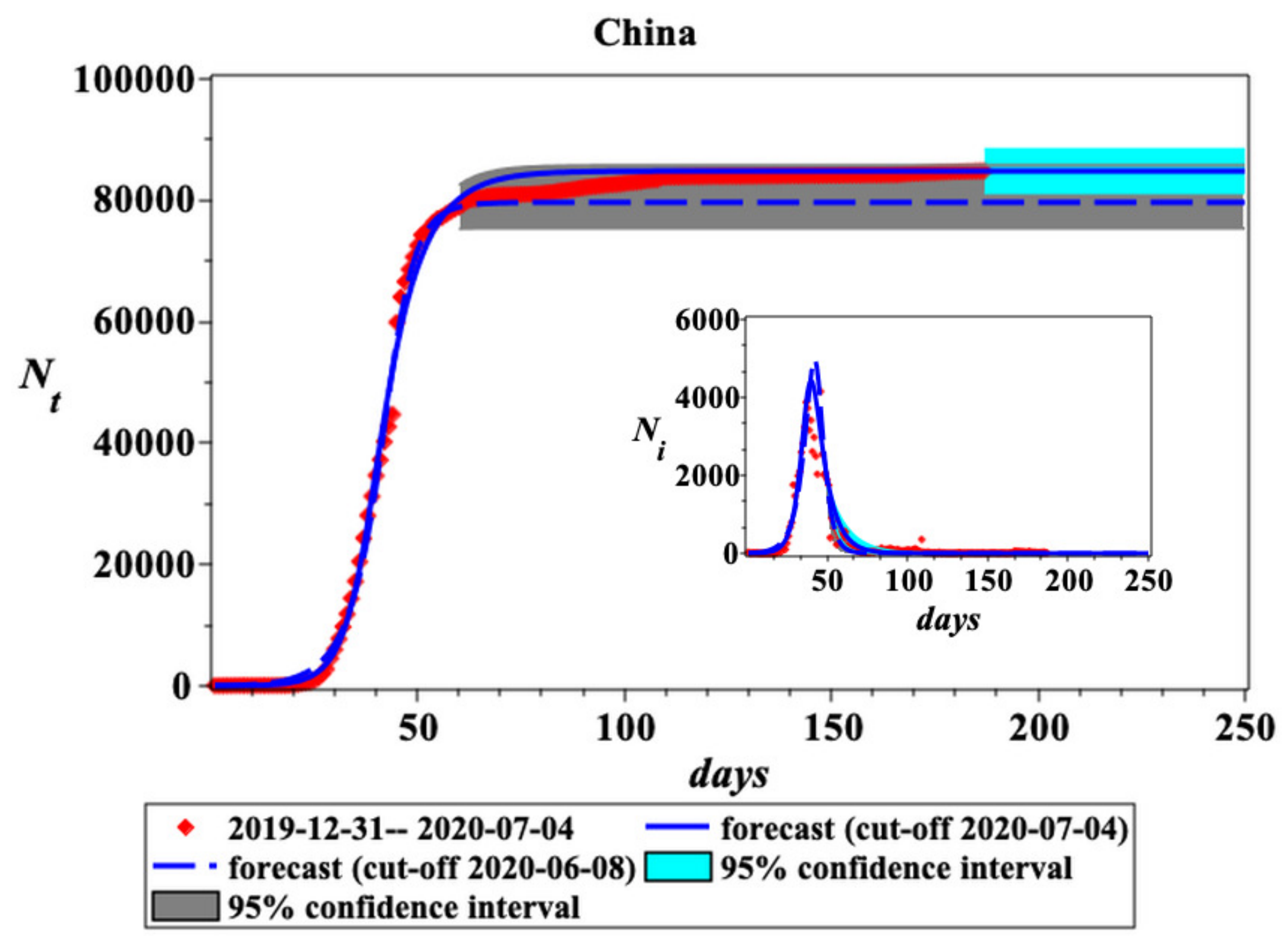}
    \end{subfigure}
       \caption{Forecast: China. Shadow bands show the $95\%$ confidence interval. Actual data: red diamonds. Insets show the number of daily new cases.}
      \label{ChinaF}
\end{figure*}

\begin{figure*}[t!]
   % \centering
    \begin{subfigure}[t]{0.8\textwidth}
        \centering
        \includegraphics[height=3.8in]{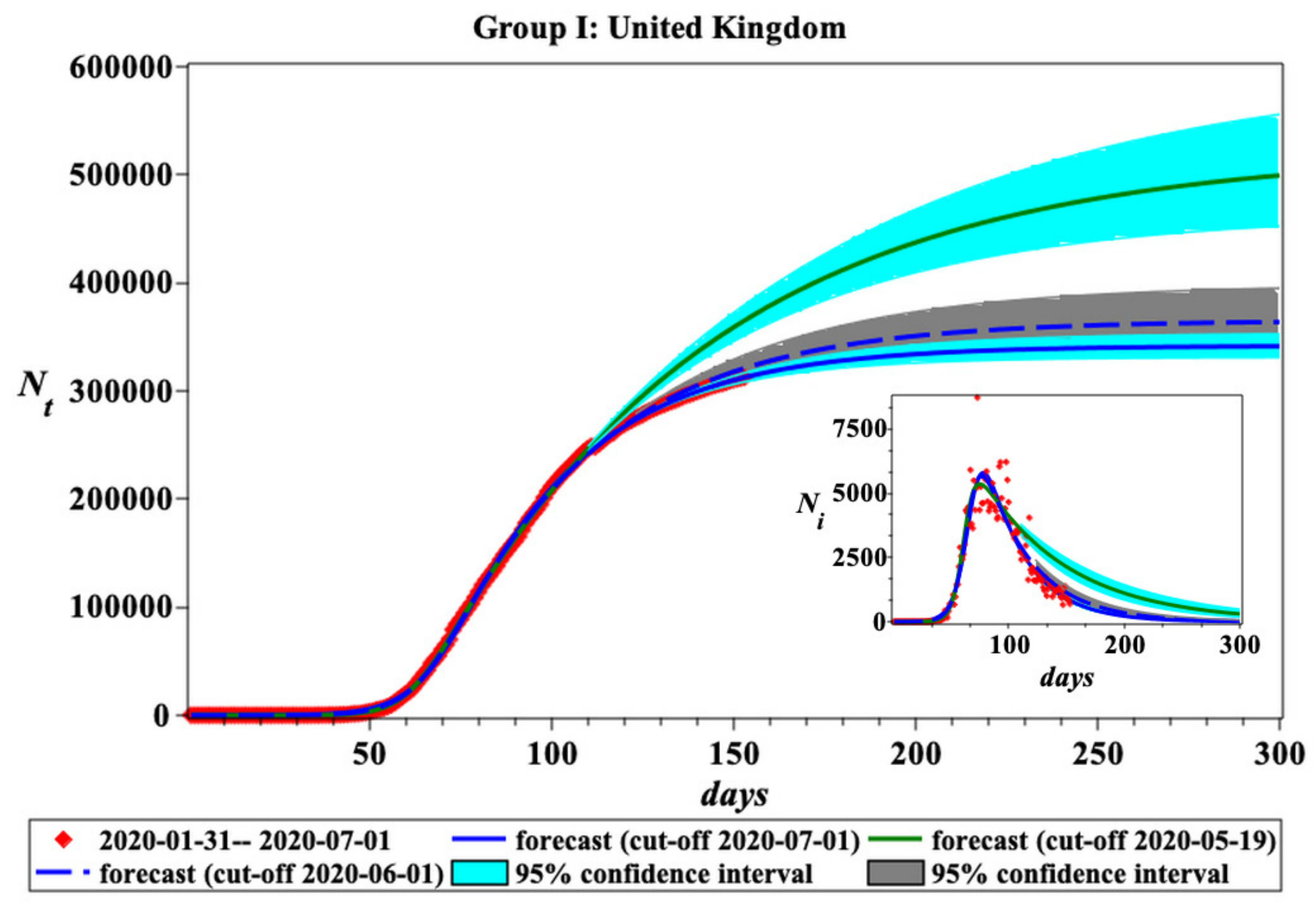}
    \end{subfigure}
       \caption{Forecast: United Kingdom. Shadow bands show the $95\%$ confidence interval. Actual data: red diamonds. Insets show the number of daily new cases.}
      \label{UKF}
\end{figure*} 
 
 For each country, the constants $A$ and $B$ are obtained from the initial and final conditions. 
 We understand the final condition as the total number of infected individuals at a cut-off date 
 within the Coronavirus Pandemic (COVID-19) database \cite{OWD}.  

 A numerical best-fit algorithm has been used to fit the model to the data by adjusting 
 parameters $\alpha$, $\kappa$, and $b$ above for each country. In principle, both the total 
 case data and the daily case data should yield the same parameter values, but the former is 
 usually less noisy. The data should begin at the first non-zero case count (i.e., $N_t > 0$), so 
 we removed any leading zeroes before this in the data for each country.

 \subsection*{COVID-19 forecast}
  
The accurate prediction of the COVID-19 pandemic spread requires the knowledge of many factors, such as human behavior, reliable tests, and government regulations. The forecast's main problem is the absence of reliable and accurate data on daily and total cases reported. Therefore the prediction should be taken with much care and, in most cases, only for a short period.
 
Our model is deterministic by nature, and therefore the prediction also is deterministic. To 
 introduce the forecast's uncertainty in case numbers, we use the residual standard deviation 
 $(S_{res})$, available in {\textit{Maple} and \textit{Mathematica}. We define the $95\%$ 
 confidence interval in 
 the trial time-domain as $\Delta N_c = 1.96S_{res}$. To obtain the plume diagram, we use the 
 probe function $N_t\pm \Delta N_c $ and fit it to the empirical 
 data in the trial interval.
 
 In Figs. \ref{ChinaF}, \ref{UKF}, we demonstrate the forecasting features of our model for China 
 and the United Kingdom. Shadow bands show the $95\%$ confidence interval. Insets show the 
 number of daily new cases. In our estimates, we assume that current interventions will 
 continue indefinitely.   
 
In the following pages, we show the algorithmically determined curves and parameters for each 
country. Though we analyzed the complete dataset for 213 countries, plus the worldwide 
aggregate, for practical purposes, we include here only countries with 50000 total cases or 
more, and also exclude some with very volatile or irregular data. As mentioned in the main text, 
some countries have begun a second wave of contagions, and they should be considered in the 
same way as Iran, Saudi {Arabia}, and the United States.
 
Countries appear in alphabetical order by country code (i.e., GBR for Great Britain, \textit{not} UK for United Kingdom), except for Worldwide data (code OWID\_WRL) which is shown last. 
\\

\begin{widetext}
\includegraphics[width=0.5\textwidth]{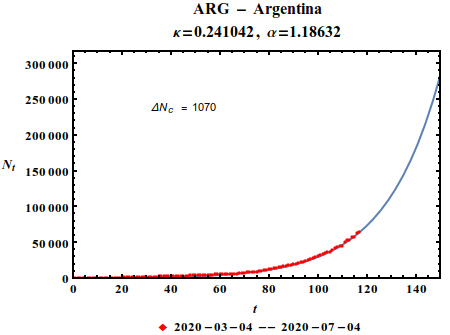} \includegraphics[width=0.5\textwidth]{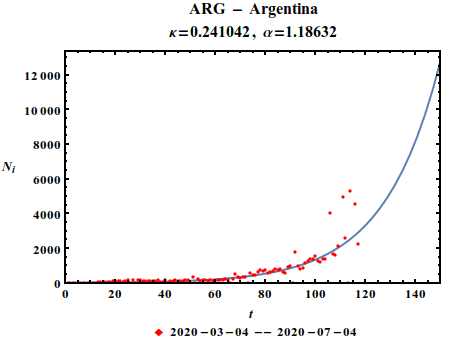} \\ 

\includegraphics[width=0.5\textwidth]{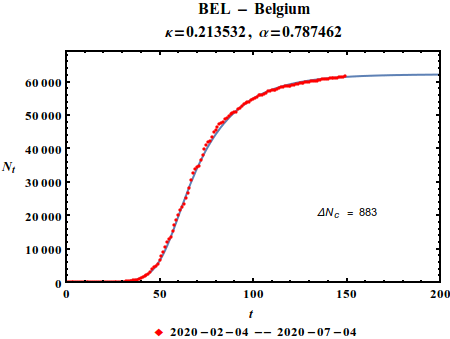} \includegraphics[width=0.5\textwidth]{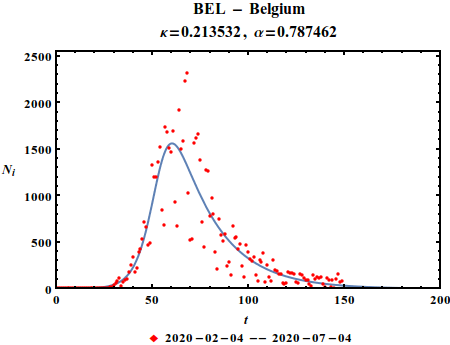} \\ 

\includegraphics[width=0.5\textwidth]{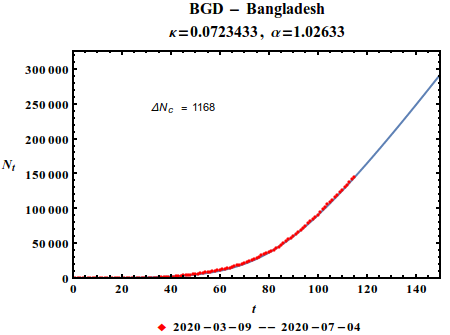} \includegraphics[width=0.5\textwidth]{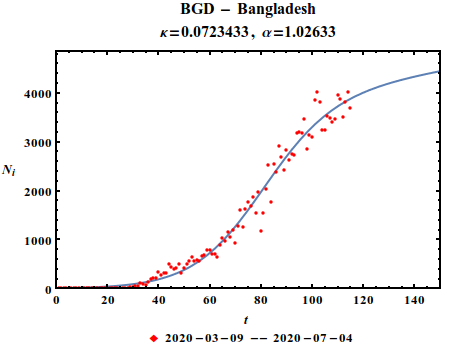} \\ 

\includegraphics[width=0.5\textwidth]{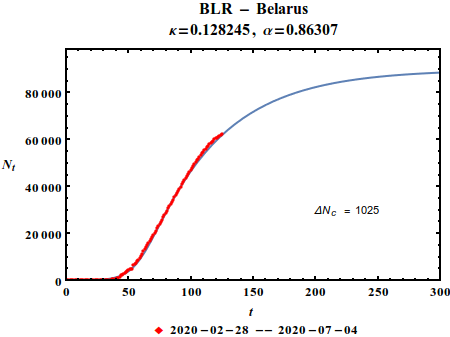} \includegraphics[width=0.5\textwidth]{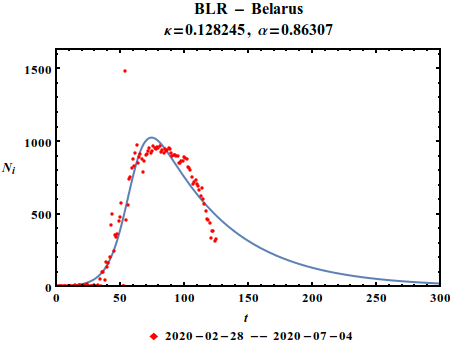} \\ 

\includegraphics[width=0.5\textwidth]{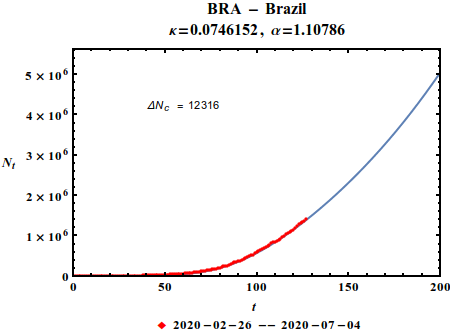} \includegraphics[width=0.5\textwidth]{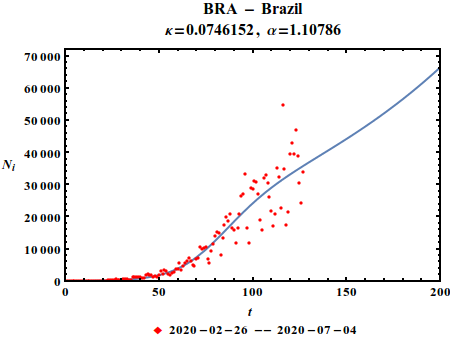} \\ 

\includegraphics[width=0.5\textwidth]{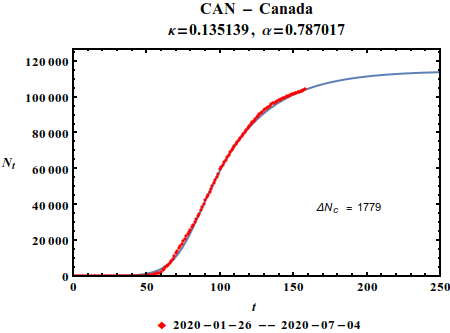} \includegraphics[width=0.5\textwidth]{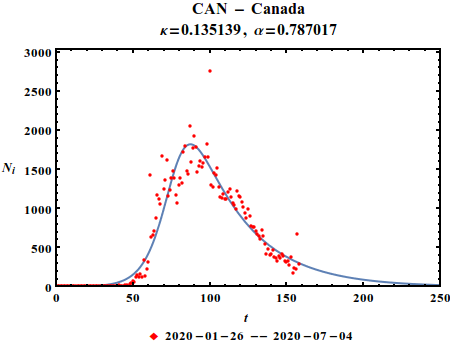} \\ 

\includegraphics[width=0.5\textwidth]{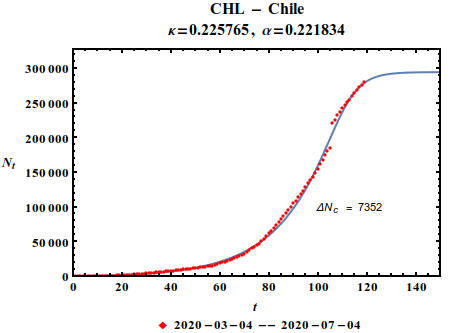} \includegraphics[width=0.5\textwidth]{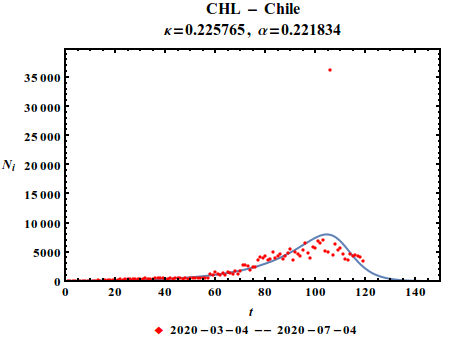} \\ 

\includegraphics[width=0.5\textwidth]{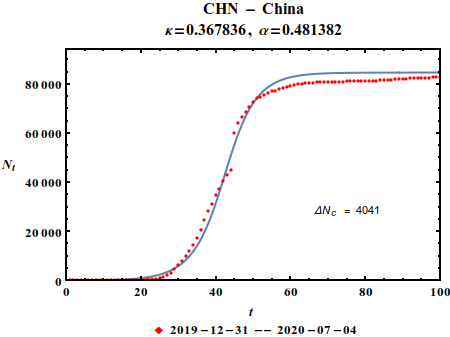} \includegraphics[width=0.5\textwidth]{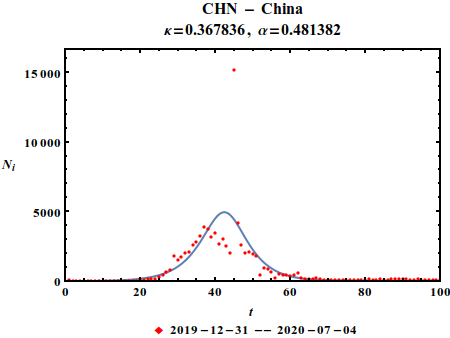} \\ 

\includegraphics[width=0.5\textwidth]{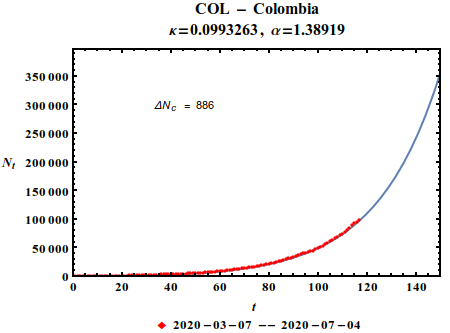} \includegraphics[width=0.5\textwidth]{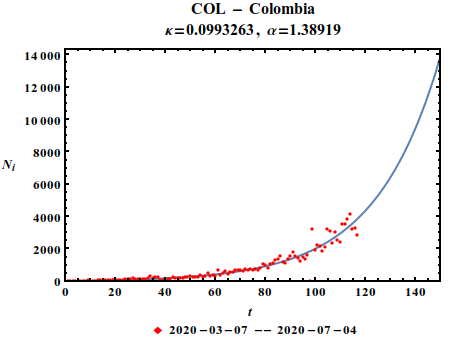} \\ 

\includegraphics[width=0.5\textwidth]{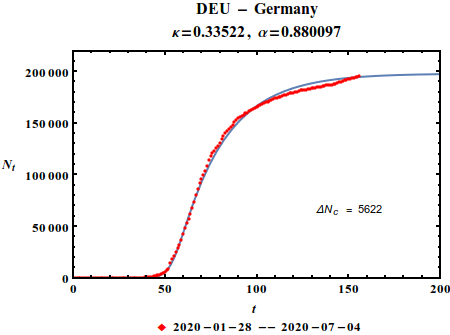} \includegraphics[width=0.5\textwidth]{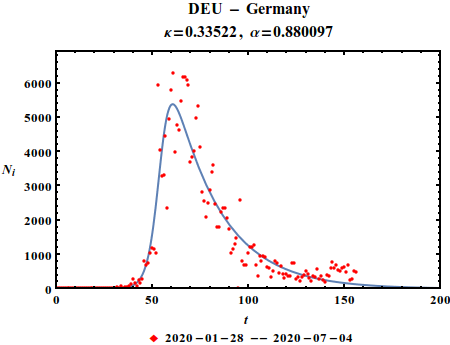} \\ 

\includegraphics[width=0.5\textwidth]{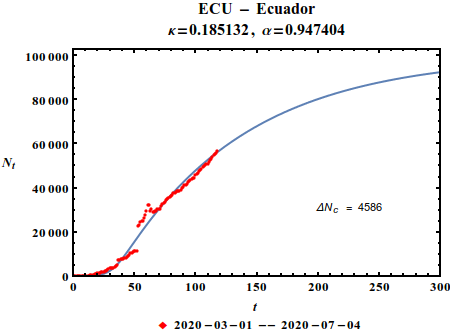} \includegraphics[width=0.5\textwidth]{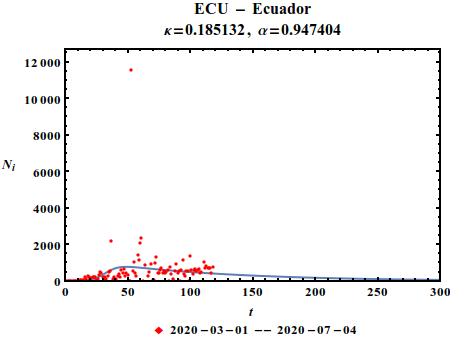} \\ 

\includegraphics[width=0.5\textwidth]{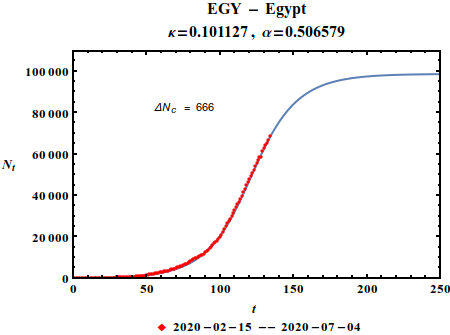} \includegraphics[width=0.5\textwidth]{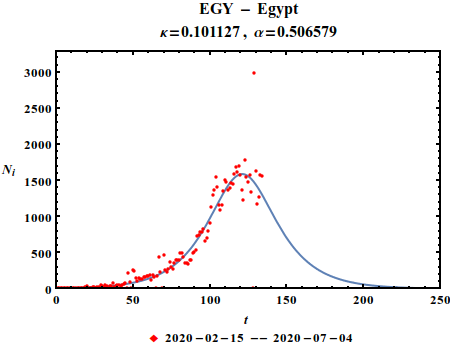} \\ 

\includegraphics[width=0.5\textwidth]{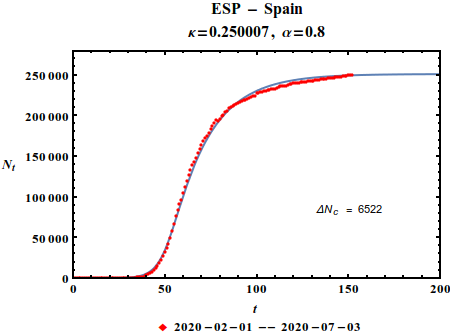} \includegraphics[width=0.5\textwidth]{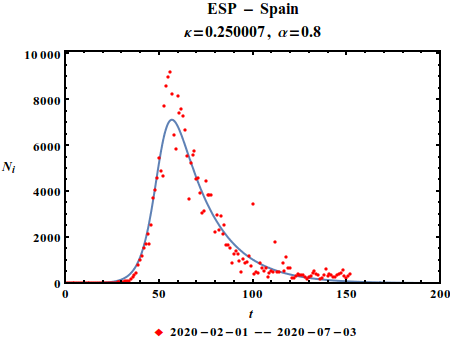} \\ 

\includegraphics[width=0.5\textwidth]{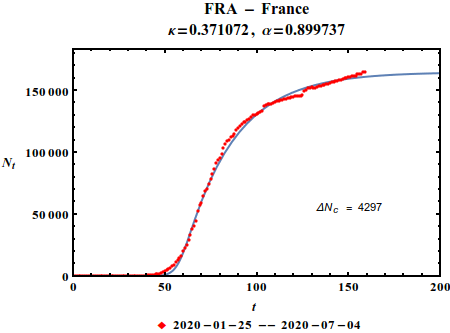} \includegraphics[width=0.5\textwidth]{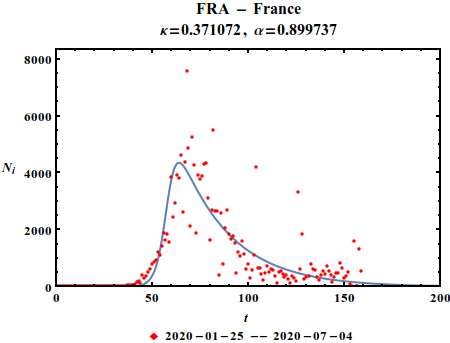} \\ 

\includegraphics[width=0.5\textwidth]{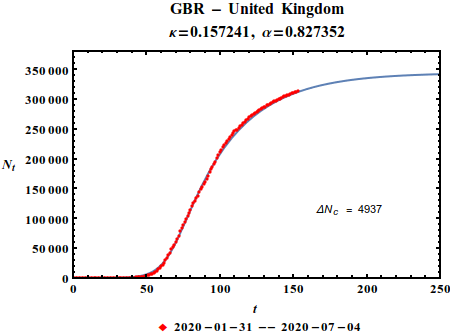} \includegraphics[width=0.5\textwidth]{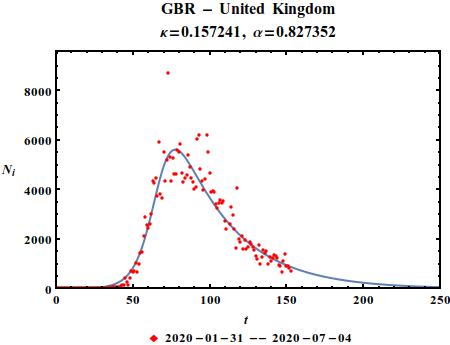} \\ 

\includegraphics[width=0.5\textwidth]{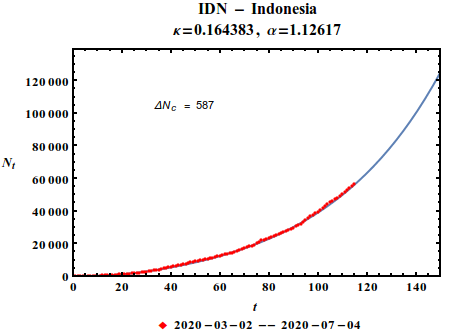} \includegraphics[width=0.5\textwidth]{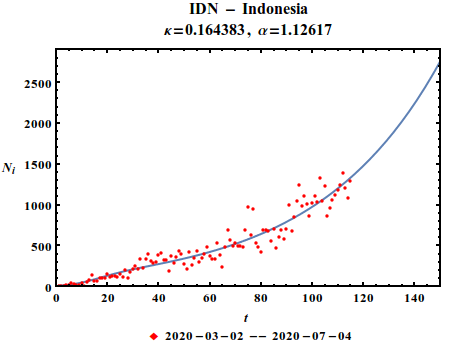} \\ 

\includegraphics[width=0.5\textwidth]{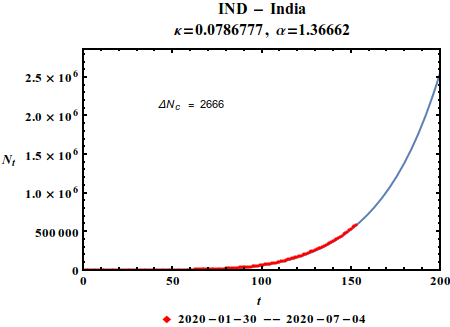} \includegraphics[width=0.5\textwidth]{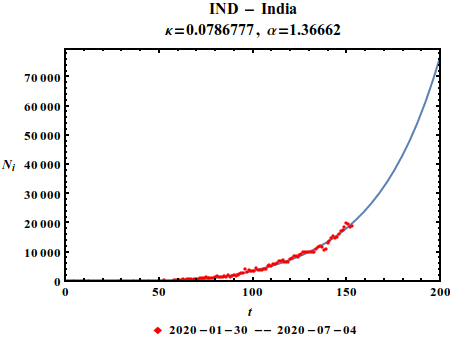} \\

\includegraphics[width=0.5\textwidth]{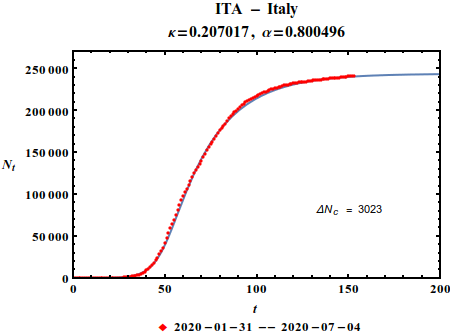} \includegraphics[width=0.5\textwidth]{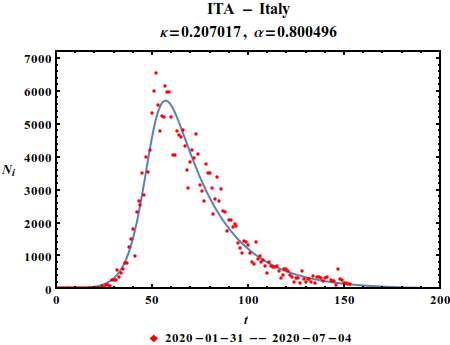} \\ 

\includegraphics[width=0.5\textwidth]{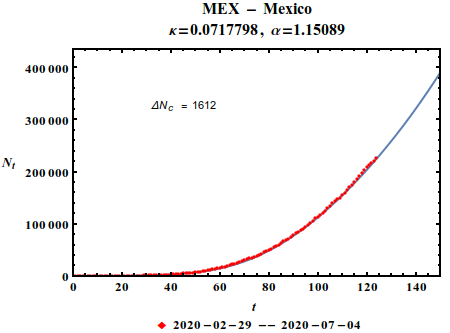} \includegraphics[width=0.5\textwidth]{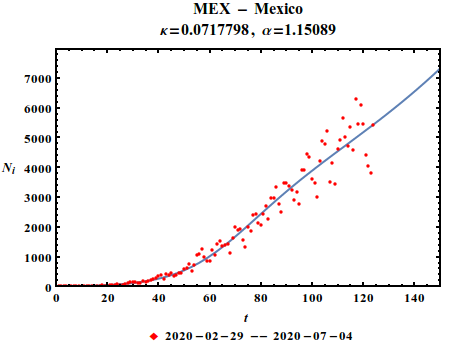} \\ 

\includegraphics[width=0.5\textwidth]{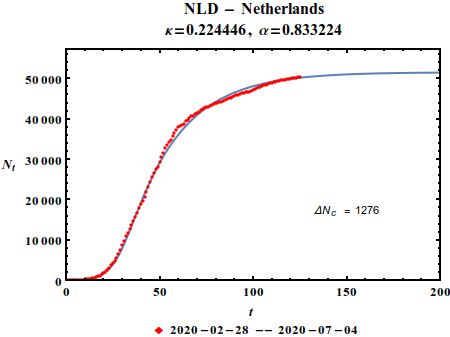} \includegraphics[width=0.5\textwidth]{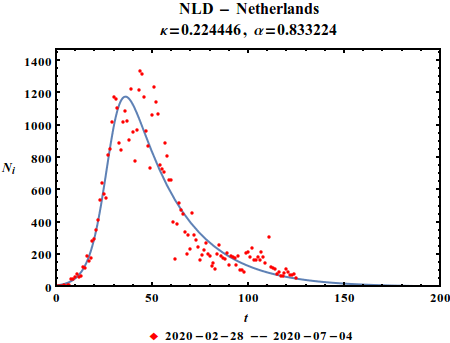} \\ 

\includegraphics[width=0.5\textwidth]{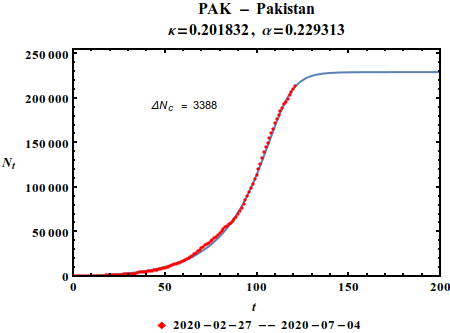} \includegraphics[width=0.5\textwidth]{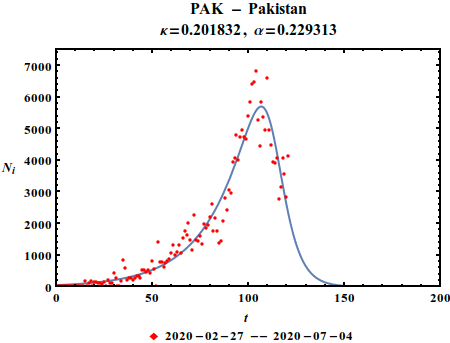} \\ 

\includegraphics[width=0.5\textwidth]{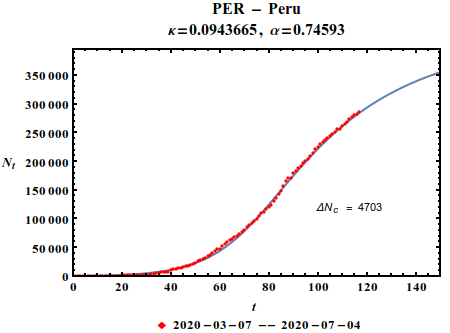} \includegraphics[width=0.5\textwidth]{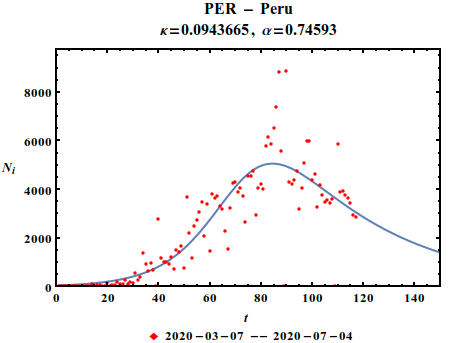} \\

\includegraphics[width=0.5\textwidth]{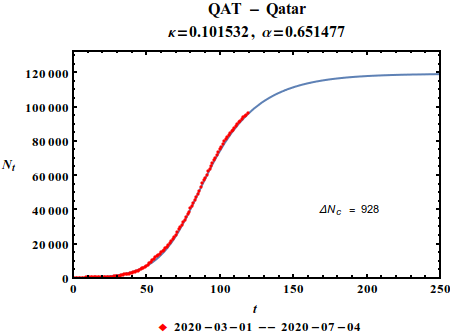} \includegraphics[width=0.5\textwidth]{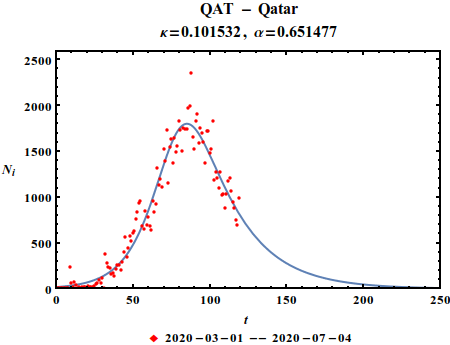} \\ 

\includegraphics[width=0.5\textwidth]{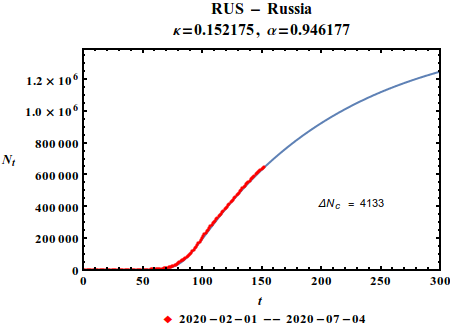} \includegraphics[width=0.5\textwidth]{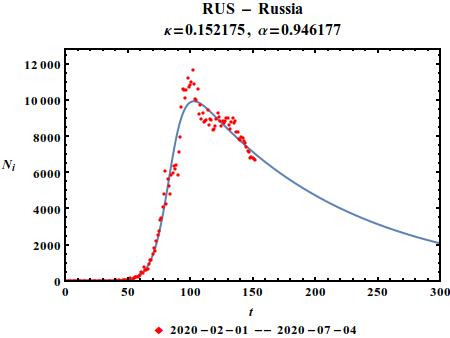} \\

\includegraphics[width=0.5\textwidth]{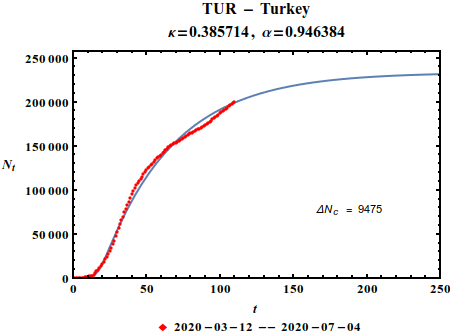} \includegraphics[width=0.5\textwidth]{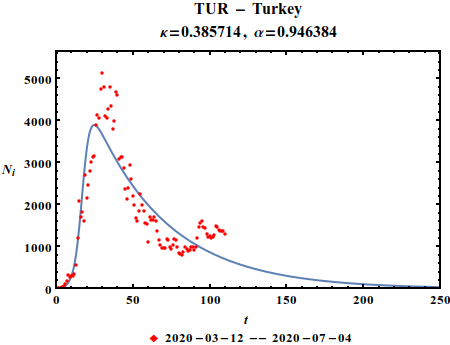} \\ 

\includegraphics[width=0.5\textwidth]{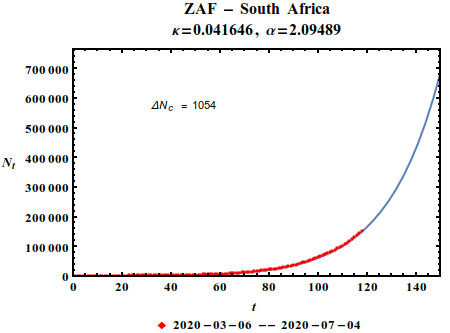} \includegraphics[width=0.5\textwidth]{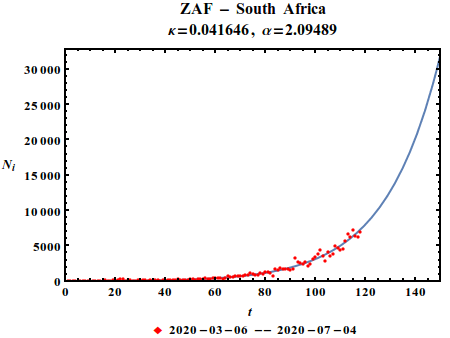} \\ 

\includegraphics[width=0.5\textwidth]{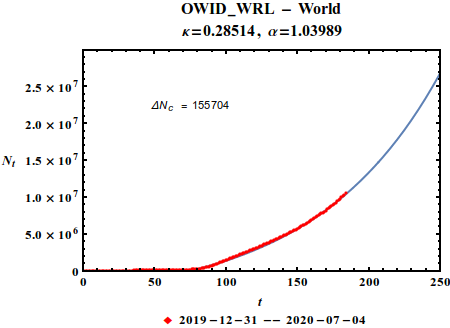} \includegraphics[width=0.5\textwidth]{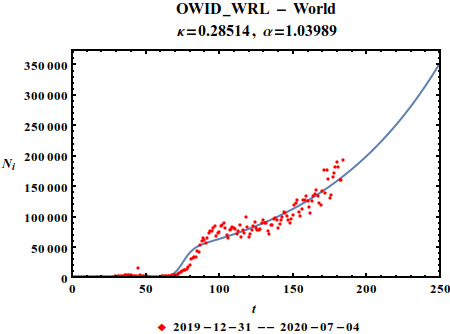} \\

\end{widetext}

\end{document}